\documentclass[12pt,preprint]{aastex}
\usepackage{morefloats}
\usepackage{color}
\shorttitle{Disk Zones}
\shortauthors{Mohanty}
\begin{document}
\def\zw{z_w }
\def\dop{\mathcal{D} }
\def\xangle{\vartheta }
\def\xanglew{\vartheta_w }
\def\Phid{{\Phi}_d }
\def\Phidx{{\Phi}_{\rm dx} }
\def\Phimx{{\Phi}_{\rm mx} }
\def\Phim{{\Phi}_{\rm m}}
\def\Phif{{\Phi}_{\rm f} }
\def\Phit{{\Phi}_t }
\def\Phiw{{\Phi}_w }
\def\Phir{{\Phi}_{\rm r} }
\def\rx{R_X }
\def\rk{R_k }
\def\del{{\bf{\nabla}} }
\def\delsq{{\nabla}^2 }
\def\vecr{{\bf{r}} }
\def\vecB{{\bf{B}} }
\def\vecBp{{\bf{B}_{\rm p}} }
\def\vecBa{{{B_{\varphi}}} }
\def\vecBr{{{B_{\varpi}}} }
\def\vecBz{{{B}_z} }
\def\malven{{\mathcal{M}}_A }
\def\rstar{R_{\ast} }
\def\rd{R_D }
\def\omstar{{\Omega}_{\ast} }
\def\omx{{\Omega}_X }
\def\ax{a_X }
\def\jbstar{\bar J_{\ast} }
\def\zmax{z_{\rm max} }
\def\deg{$^{\circ}$ }
\def\erhat{{\^{e}}$_{\varpi}$}
\def\eahat{{\^{e}}$_{\varphi}$}
\def\ezhat{{\^{e}}$_z$}
\def\mstar{M_{\ast} }
\def\rstar{R_{\ast} }
\def\lstar{L_{\ast} }
\def\msun{M_{\odot} }
\def\rsun{R_{\odot} }
\def\lsun{L_{\odot} }
\def\mdotd{\dot M_D }
\def\mdot{\dot M }
\def\rhos{{\rho}_0 }
\def\vs{v_0 }
\def\rhoa{{\rho}_{Am} }
\def\va{v_{Am} }
\def\bs{B_0 }
\def\ba{B_{Am} }
\def\rs{r_0 }
\def\ra{r_{Am} }
\def\rr{r_{A2} }
\def\vp{{\mathcal{V}}_0 }
\def\h2{{\rm H}_2 }
\def\met{{\rm M}^{+} }
\def\avz{\langle Z \rangle }
\def\gz{{\rm gr}\avz }
\def\gn{{\rm gr}0 }
\def\gneg{{\rm gr}^{-} }
\def\nh2{n_{\h2} }
\def\ne{n_e }
\def\ni{n_i }
\def\ngz{n_{\gz} }
\def\ngn{n_{\gn} }
\def\ngneg{n_{\gneg} }
\def\xe{x_e }
\def\xi{x_i }
\def\xgz{x_{\gz} }
\def\xgneg{x_{\gneg} }
\def\aie{\alpha_{i,e} }
\def\agze{\alpha_{e,\gz} }
\def\agzi{\alpha_{i,\gz} }
\def\agne{\alpha_{e,\gn} }
\def\agni{\alpha_{i,\gn} }
\def\agnq{\alpha_{q,\gn} }
\def\zetax{\zeta_X }
\def\col{N_{\perp} }
\def\baraie{{\bar{\alpha}}_{i,e} }
\def\xgcross{x_{\rm gr}^{\ast} }

\title{Dead, Undead and Zombie Zones in Protostellar Disks \\ as a Function of Stellar Mass}
\author{Subhanjoy Mohanty\altaffilmark{1}, Barbara Ercolano\altaffilmark{2,3}, Neal J.\ Turner\altaffilmark{4}}
\altaffiltext{1}{Imperial College London, 1010 Blackett Lab., Prince Consort Road, London SW7 2AZ, UK.  s.mohanty@imperial.ac.uk}  
\altaffiltext{2}{Universit\"ats-Sternwarte M\"unchen, Scheinerstr. 1, 81679 M\"unchen, Germany}  
\altaffiltext{3}{Cluster of Excellence Origin and Structure of the Universe, Boltzmannstr.2, 85748 Garching, Germany.  ercolano@usm.lmu.de}  
\altaffiltext{4}{Jet Propulsion Laboratory, California Institute of Technology, Pasadena, CA 91109, USA.  neal.turner@jpl.nasa.gov}  

\begin{abstract}
  We investigate the viability of the magnetorotational instability
  (MRI) in X-ray ionized viscous accretion disks around both
  solar-type stars and very low mass stars.  In particular, we
  determine the disk regions where the MRI can be shut off either by
  Ohmic resistivity (the so-called Dead and Undead Zones) or by
  ampipolar diffusion (a region we term the Zombie Zone).  We consider
  2 stellar masses: $\mstar = 0.7 \msun$ and 0.1$\msun$. In each case,
  we assume that: the disk surface density profile is that of a scaled
  Minimum Mass Solar Nebula, with $M_{disk}/M_{\ast}$ = 0.01 as
  suggested by current data; disk ionisation is driven primarily by
  stellar X-rays, complemented by cosmic rays and radionuclides; and
  the stellar X-ray luminosity scales with bolometric luminosity as
  $L_X/\lstar \approx 10^{-3.5}$, as observed.  Ionization rates are
  calculated with the {\sc moccasin} Monte Carlo X-ray transport code,
  and ionisation balance determined using a simplified chemical
  network, including well-mixed 0.1 $\mu$m grains at various levels of
  depletion.  We find that {\it (1)} ambipolar diffusion is the
  primary factor controlling MRI activity in disks around both
  solar-type and very low mass classical T Tauri stars.  Assuming that
  the MRI yields the maximum possible field strength at each radius,
  we further find that: {\it (2)} the MRI-active layer constitutes
  only $\sim$5--10\% of the total disk mass; {\it (3)} the accretion
  rate ($\mdot$) varies radially in both magnitude and {\it sign}
  (inward or outward), implying time-variable accretion as well as the
  creation of disk gaps and overdensities, with consequences for
  planet formation and migration; {\it (4)} achieving the empirical
  accretion rates in solar-type and very low mass stars requires a
  depletion of well-mixed small grains (via grain growth and/or
  settling) by a factor of 10--1000 relative to the standard
  dust-to-gas mass ratio of $10^{-2}$; and {\it (5)} the current
  non-detection of polarized emission from field-aligned grains in the
  outer disk regions is consistent with active MRI at those radii.
%; and {\it (4)} the precise magnitude of $\mdot$ is also sensitive to the assumed chemical network.  
\end{abstract}

\section{Introduction}

The inward migration of material in protoplanetary disks requires a source of anomalous viscosity; microscopic gas viscosity is far too inefficient to explain the observed accretion rates and disk lifetimes.  The best current candidate is turbulent viscosity generated by the magneto-rotational instability (MRI), which arises due to Keplerian shearing of weak magnetic fields embedded in a partially ionized disk \citep{1991ApJ...376..214B}.  Assuming ionization due to cosmic rays, Gammie (1996) argued that MRI-driven accretion in protoplanetary disks is likely to be ``layered'': near the mid-plane in the inner disk, where column densities are too high for cosmic rays to penetrate, Ohmic resistivity quenches the MRI and forms a Dead Zone, while active MRI proceeds in the surface layers of the inner disk as well as throughout the outer disk, where low densities permit sufficiently high fractional ionization for good gas-field coupling.  \citet[hereafter IG99]{1999ApJ...518..848I} subsequently showed that ionization by stellar X-rays overwhelms that by cosmic rays, but this does not change the basic paradigm of an Ohmic-dominated Dead Zone hemmed in by active disk regions.  

It has become increasingly clear, though, that this picture is overly
simplified, because Hall and ambipolar diffusivities cannot be ignored
under the conditions prevalent in these disks
\citep[e.g.,][]{1998ApJ...501..758H, 1999MNRAS.307..849W,
  2001ApJ...552..235B, 2002ApJ...577..534S, 2004MNRAS.348..355K,
  2007ApSS.311...35W}.  Specifically, active MRI in this scenario is
only possible where the density is low enough to allow adequate
fractional ionization (i.e., outside the Dead Zone); however, it is
precisely at these low densities that either electrons, or both
electrons and ions, can become decoupled from the neutrals, promoting
Hall diffusion in the first instance and ambipolar in the second.  The
non-dissipative nature of the Hall term means that it may enhance,
suppress or leave unaffected the MRI efficiency, depending on the
precise environment and magnetic field geometry
\citep{1999MNRAS.307..849W, 2002ApJ...577..534S, 2011arXiv1103.3562W}.
The ambipolar term, though, {\it is} dissipative, and can smother the
MRI by decoupling the field entirely from the predominantly neutral
gas.  Dust grains do not change these conclusions, but only add to the
dangers facing the MRI, by soaking up electrons and facilitating
ion-electron recombinations on their surfaces, thus greatly reducing
the ionization fraction and increasing the resistivities
\citep[e.g.,][hereafter IN06]{2000ApJ...543..486S,
  2006AA...445..205I}.

In a seminal analysis, \citet{2011ApJ...736..144B} have investigated in detail the effects of ambipolar diffusion on the non-linear evolution of the MRI, through local 3D shearing-box simulations.  Their calculations are in the ``strong coupling'' limit (wherein the ion density is negligible compared to the neutral density, and the electron recombination time is much smaller than the orbital period), and explicitly account for the non-conservation of ion densities (which depend instead on the chemical equilibrium).  Their analysis yields a quantitative general criterion for evaluating whether or not ambipolar diffusion can extinguish the MRI.        

Following this, \citet[hereafter B11]{2011ApJ...739...50B} showed that the ``strong coupling'' limit does invariably hold in protoplanetary disks.  Applying the above criterion to X-ray irradiated disks around solar-type stars, he further finds that ambipolar diffusion strongly determines the MRI efficiency in such disks, and, in the presence of small (0.1\,$\mu$m) grains, drives the attendant accretion rate below the observed rates in these stars by an order of magnitude or more.  In an elegant and complementary semi-analytic study, \citet[hereafter PC11]{2011ApJ...727....2P} find an even graver suppression of the MRI with polycyclic aromatic hydrocarbon (PAH) grains.  These somber implications for the viability of the MRI are alleviated somewhat by \cite{2011ApJ...735....8P}, who contend that stellar FUV photons produce much more ionized surface plasmas than X-rays, yielding efficient MRI and adequate accretion rates beyond $\sim$10\,AU in solar-type disks, in spite of PAHs.  Even in this case, however, the inferred accretion in the inner disk remains much weaker than observed.  

Our goals in this paper are twofold.  First, we wish to extend the above studies of X-ray irradiated disks to very low mass stars ($\sim$0.1\,$\msun$, i.e., mid-M spectral types).  It is only by expanding the parameter space in this fashion that we can hope to fully understand the feasibility and importance of the MRI for driving viscous accretion.  For instance, it has been suggested that the disks girdling such stars may be far more active than their more massive counterparts around solar-type stars, because their lower column densities enable ionizing photons to penetrate much deeper \citep[e.g.,][]{2006ApJ...648..484H}.  In that case, perhaps the problems, alluded to above, that plague the MRI in more massive disks are not as severe for very low mass stars?  However, as discussed, lower densities also enhance ambipolar diffusion, so it is a priori not at all clear if this hypothesis is valid (as we will see, it is not).  Moreover, there are now burgeoning observations of planets down to terrestrial masses orbiting low-mass M dwarfs \citep[e.g.,][]{2011arXiv1111.5019B}, and understanding their formation and migration is inextricably linked to deciphering the viscosity and accretion properties of their natal disks.  

Second, we wish to implement certain important improvements over the
previous analyses.  In particular, we undertake a more sophisticated
treatment of the X-ray ionization, including Compton scattering, with
the 3-D Monte Carlo photoionization and dust radiative transfer code
{\sc moccasin}.  We also derive accretion rates via a more generalized
equation, and show in the process that radial gradients have some
important implications (e.g., time-dependent accretion rates, and the formation of disk gaps and overdensities).  Furthermore, we examine the effects of
various levels of grain depletion, intermediate between the limiting
cases of a standard ISM dust abundance and no dust at all.  Finally,
for both solar-type and very low mass stars, we compute the detailed
chemical, ionization and MRI properties over the entire disk from 0.1
to 100\,AU, instead of at a selected few radial locations.  Save these
changes (and some differences in the chemical network, discussed at
the appropriate junctures), our study is analogous to that of B11.

The layout of the paper is as follows.  We outline the criteria for active MRI in \S2, and discuss the conditions for determining the relevant field strengths and accretion rates in \S3.  In \S4, we describe the adopted disk physical properties -- the disk structure (\S4.1), chemical and ionization balance (\S4.2), and resulting resistivities (\S4.3) -- used to evaluate the MRI criteria given in \S2.  The specific stellar cases we examine are summarized in \S5.  Our results are presented in \S6, and conclusions and directions for future work laid out in \S7.

\section{Conditions for Magnetically Driven Disk Accretion}
\subsection{Dead and Undead Zones}
Magnetic stresses can transport angular momentum in the disk, and thus
drive disk accretion, only if the gas is sufficiently ionized to
couple to the field, i.e., if gas motions can generate magnetic
disturbances faster than these can diffuse away due to the finite
resistivity $\eta$.  In a Keplerian disk, it is fundamentally the
orbital shear that creates the stresses, so the relevant minimum
timescale for field generation is the orbital period, $\sim 1/\Omega$.
For {\it large-scale} toroidal fields produced by such shearing of a
weak seed radial field, the characteristic length-scale for field
variations is of order the radial distance to the star, $\sim
v_K/\Omega$, where $v_K$ is the Keplerian velocity.  The dissipation
timescale is thus $\sim (v_K/\Omega)^2/\eta_O$, if Ohmic resistivity
$\eta_O$ provides the primary diffusion mechanism.  The condition for
field regeneration to prevail over dissipation is then given by the
Reynolds number criterion \citep{2008ApJ...679L.131T}:
$$ \Lambda_K \equiv \frac{v_K^2}{\eta_O \Omega} > 10 \eqno(1) $$
with the lower limit chosen to give toroidal fields 10 times stronger
than the initial seed.

For {\it local} tangled fields generated by MRI-driven turbulence, the height of the thin disk sets an upper limit on the wavelength of vertical MRI modes, and thereby on the dissipation timescale, so it is the vertical direction we must consider.  For a vertical mode with wavenumber $k$, the Ohmic dissipation rate is $\sim k^2 \eta_O$ while the growth rate is $k\, v_{{\mathcal A}}\,$, where $v_{{\mathcal A}}$ is the local Alfv\'{e}n velocity.  The maximum growth rate is again the orbital frequency $\sim \Omega$; the fastest growing mode thus has a wavenumber $k = \Omega/v_{{\mathcal A}}$.  Demanding that its growth rate exceed its dissipation rate then yields the Elsasser number criterion:
$$ \Lambda_{\mathcal A} \equiv \frac{v_{{\mathcal A}}^2}{\eta_O \Omega} > 1 \eqno(2) $$
whether the net background field is vertical, toroidal or zero \citep{2002ApJ...577..534S}.  As noted by \citet{2007ApJ...659..729T}, the above Elsasser number criterion captures MRI activity more accurately than an analogous version with $v_{{\mathcal A}}$ replaced by the sound speed $c_s$ \citep[e.g., as adopted by][]{2000ApJ...530..464F, 2007ApSS.311...35W}.  The specification of the Alfv\'{e}n velocity $v_{{\mathcal A}}[r] \equiv B[r]/\sqrt{4\pi \rho[r]}$ is described in \S3.  For now, note two consequences of the requirement (see \S2.2 below) that the gas pressure dominate over the magnetic pressure in the disk.  First, this implies $v_{{\mathcal A}}/c_s \ll 1$, which guarantees that the wavelength $\sim v_{{\mathcal A}}/\Omega$ of the fastest growing mode near the midplane is indeed much less than the disk scale height $z_H  \sim c_s/\Omega$. 

Second, combining the requirement $v_{{\mathcal A}}/c_s \ll 1$ with
the thin disk condition $v_K \gg c_s$ implies $v_K \ggg v_{{\mathcal
    A}}$.  Hence, of the two criteria (1) and (2), the first is the
less restrictive, in that it allows field generation at much higher
resistivities than the second.  Thus the two divide the disk into 3
parts.  In the active zone, both (1) and (2) are satisfied, and
efficient magnetically-driven angular momentum transport can occur in
spite of Ohmic resistivity.  In the Undead Zone, the MRI is quenched,
but orbital shear can still generate large-scale toroidal fields out
of radial ones siphoned off from neighbouring active layers (if any
exist).  In the Dead Zone, both processes are quenched by Ohmic
dissipation, the gas is almost entirely decoupled from the field, and
magnetic torques cannot drive angular momentum transport.

\subsection{Ambipolar Diffusion and the Zombie Zone}
The criteria in equations (1) and (2) have been formulated in terms of the Ohmic resistivity.  However, Hall or ambipolar diffusivities can be expected to dominate over Ohmic in various parts of the disk.  What are the criteria for MRI under such conditions?

Equation (2) simply expresses the condition that, for robust MRI in the Ohmic domain, the ratio of the inductive to the Ohmic resistive term in the induction equation must exceed unity.  When Hall or ambipolar diffusivity is the major resistive term instead, the analogous condition is naively obtained by replacing $\eta_O$ in equation (2) by $|\eta_H|$ (Hall) or $\eta_A$ (ambipolar) (see \S4.3).  However, local 3-D MHD simulations by \citet{2002ApJ...577..534S} show that, while the Hall term can affect the saturation level of the Maxwell stress by a factor of a few, it hardly changes the critical condition for strong MRI: this remains the Ohmic Elsasser criterion equation (2), regardless of the size of the Hall term or the strength and geometry of the field.  {\it Hence we use equation (2) in the Hall domain as well} (i.e., calculate the Elsasser number using $\eta_O$ even when $|\eta_H|$ dominates, thereby ignoring Hall effects altogether).  \citet{2011arXiv1103.3562W} show that the aforementioned simulations do not probe all the pertinent parameter space, and perform a linear analysis that suggests Hall diffusion may have a powerful effect on the MRI in a previously unexplored parameter regime that is nevertheless very germane to protoplanetary disks.  We discuss their results briefly in \S7; however, in the absence of non-linear simulations here, we neglect this added complication for the present analysis.  

In the ambipolar regime, on the other hand,
\citet{1999MNRAS.307..849W} argues that it {\it is} appropriate to
substitute $\eta_A$ for $\eta_O$ in equation (2), leading to the
ambipolar Elsasser number condition
$$ Am \equiv \frac{v_{{\mathcal A}}^2}{\eta_A \Omega} > 1 \eqno(3) $$ 
for the MRI to exist here.  Moreover, if electrons and ions are the
only charged species, this reduces to $\gamma_i\rho_i/\Omega > 1$ (see
\S4.3), where $\gamma_i$ is the neutral-ion collisional drag
coefficient and $\rho_i$ the ion density; this implies that the MRI
can occur under these circumstances if a neutral particle collides
with at least one ion on average per orbital period.  This condition
has been invoked by several authors
\citep[e.g.,][]{2010ApJ...708..188T} in ambipolar-dominated disk
regions.

Until recently, the only existing simulations of MRI with ambipolar diffusion, by \citet{1998ApJ...501..758H}, indicated that the above criterion was actually too liberal.  Using 3-D MHD local shearing box simulations, based on an idealised 2-component fluid (ions + neutrals), these authors found that {\it efficient} angular momentum transport by the MRI requires neutral-ion collisions to be at least a hundred times more frequent, i.e., $\gamma_i \rho_i/\Omega \gtrsim 100$.  For $\gamma_i \rho_i/\Omega \sim 1$, the linear instability is still present, but the nonlinear evolution of the MRI is controlled primarily by the ion density, and the total angular momentum transport becomes proportional to the ionisation fraction.  In other words, when $\gamma_i \rho_i/\Omega \sim 1$, the MRI is alive only in the ions, and effectively dead in the neutrals \citep[with the neutrals only providing a drag that reduces the angular momentum transport in the ions;][]{1998ApJ...501..758H}.  

However, very recent simulations by \citet{2011ApJ...736..144B},  based on a multi-component fluid including grains (very similar to the system adopted here; \S\S4.2, 4.3), suggest that the above results are overly pessimistic.  With the ratio of the inductive to the ambipolar term now given by the full general value $v_{{\mathcal A}}^2/\eta_A \Omega$ (instead of the reduced value $\gamma_i \rho_i/\Omega$), these authors find that the MRI can be sustained at {\it any} value of $Am$, {\it provided} the field is sufficiently weak.  Specifically, the MRI can operate as long as the plasma $\beta$-parameter, $\beta \equiv P_{gas}/P_B$, satisfies
$$\beta > \beta_{\rm min} \eqno(4a) $$    
where the limiting value is given by
$$ \beta_{\rm min} = \left[\left(\frac{50}{Am^{1.2}}\right)^2 + \left(\frac{8}{Am^{0.3}} + 1\right)^2 \right]^{1/2}\eqno(4b) $$
with $Am$ given by equation (3) and $P_B = B^2/8\pi$.  We follow B11 in adopting the above condition for MRI in ambipolar-dominated regions.  If equation (4) is {\it not} satisfied in these regions, then the MRI still drives angular momentum transport within the {\it charged} species that are tied to the field, but is quenched in the {\it neutral bulk of the disk}, since the neutrals no longer couple to the field through collisions with charged particles.  As such, the net transport of angular momentum via the MRI becomes negligible.  We call this ``marginally alive, effectively dead'' region the Zombie Zone.  

Finally, note that $\beta_{\rm min}$ approaches $(50/Am^{1.2})$ for $Am \lesssim 1$, and asymptotes to 1 from above as $Am \rightarrow \infty$.  Thus the above condition for sustainable MRI demands that the gas pressure dominate over the magnetic pressure in the disk, as stated earlier in \S2.1.   

To summarize, our conditions for active MRI are: $\Lambda_K > 10$, $\Lambda_{\mathcal A} > 1$ and $\beta > \beta_{\rm min}$.  Note that $\beta_{\rm min}$ is a function of $Am$, which itself depends on the field strength through both the Alfv\'{e}n velocity and $\eta_A$ (see \S4.3).  Similarly, the extent of the Dead Zone also depends on the magnetic field via the Alfv\'{e}n velocity.  We describe below our method for determining the field strength, based on considerations of the accretion rate.    

\section{Accretion Rate}
For radial angular momentum transport through magnetic torques, the accretion rate (positive in the inward direction) at any radius is given by \citep{2007ApSS.311...35W}:
$$ \mdot = \frac{2}{v_K}\, \frac{\partial}{\partial r}\left(r^2 h \langle -B_r B_{\phi} \rangle \right) \eqno(5a)$$
where $v_K = r \Omega$ is the Keplerian velocity, and we have defined
$$ \langle -B_r B_{\phi}\rangle \equiv -\frac{1}{2h}\int_{-s}^s B_r B_{\phi} dz \eqno(5b)$$ 
with $\pm s$ being the upper and lower surfaces of the disk, and $2h$ the total thickness of the MRI-active layers (counting both above and below the disk midplane).  Simulations of the MRI indicate that $B_{\rm rms}^2 \sim 4\langle -B_r B_{\phi}\rangle$ \citep{2004ApJ...605..321S}; we may therefore write:
$$ \mdot = \frac{1}{2\, r\Omega}\, \frac{\partial}{\partial r}\left(r^2 h B_{\rm rms}^2 \right) \eqno(6)$$
{\it If} $\mdot$ is constant with radius, then the above reduces to (integrating over $r$, and dropping the `rms' subscript on $B$ for brevity) $\mdot = h B^2/4\,\Omega$.  This is the formula used by B11 to calculate the accretion rate.  Notice that this automatically implies that $hB^2 \propto r^{-3/2}$ for self-consistency with the assumption of a radially invariant $\mdot$.  As we will show, however, there is no fundamental reason for $hB^2$ to follow precisely this radial dependence.  In particular, if this quantity is some arbitrary power-law with radius at a given location, i.e., $hB^2 \propto r^{-n}$, then the general solution to the accretion rate there is $\mdot = (2-n)\,hB^2/2\,\Omega$.  For $n = 3/2$, this reduces to the constant $\mdot$ solution above, but for other $n$ the accretion is radially variable; moreover, $\mdot$ vanishes or even becomes negative (outward radial flow) for $n \geq 2$.  

This brings us to the question of how to set the magnetic field strength.  B11 determines $B$ by assuming that it is the maximum allowed by the condition in equation (4), $\beta = \beta_{\rm min}$.  He justifies this on the basis that it allows one to calculate the highest possible accretion rate, under the assumption of a radially constant $\mdot$ = $h B^2/4\,\Omega$.  As we will show, however, $\mdot$ is {\it not} constant when $B$ is chosen this way.  Nevertheless, we can still support such a choice of $B$ with a more physically motivated justification: that {\it the MRI is maximally efficient, generating the strongest field possible that still allows the MRI to operate}.         

Our procedure therefore is as follows.  Assuming the magnetic field to
be vertically constant, we loop through a range of field strengths at
every location to find the permitted $B_{\rm max}$ (corresponding to
$\beta_{\rm min}$) at every radius.  More precisely, for every value
of $B$ (which also determines the Alfv\'{e}n velocity and the
ambipolar resistivity $\eta_A$), we calculate the thickness $h$ of the
active layer given our criteria for MRI in \S2, and choose the $B$ for
which the quantity $hB^2$ -- the height-integrated stress -- is maximized.  This is because maximizing
$B$ alone formally leads to an infinitesimal $h$, implying $\mdot
\rightarrow 0$.  However, as we will see, $hB^2$ increases in step
with $B$ until a maximum value, and then falls off extremely rapidly
to zero as $B$ is increased further; thus, for all practical purposes,
the two maxima coincide, i.e., $(hB^2)_{\rm max}$ occurs essentially
at (very slightly lower than) $B \approx B_{\rm max}$.  We then insert this $(hB^2)_{\rm max}$
into the general solution, equation [6], to calculate $\mdot$ at every
radius.  To summarize, we set the field strength to maximize the
height-integrated stress in the same way as B11 does, but our $\mdot$
differ from his since we allow for radial stress gradients.

To carry out these calculations, we must specify the disk structure, ionization balance and chemistry, and the attendant resistivities.  These are described below.  

\section{Disk Structure, Ionization Chemistry and Resistivity}

\subsection{Disk Structure}
We consider a star of mass $\mstar$, radius $\rstar$ and bolometric
luminosity $\lstar$, surrounded by a non-self-gravitating Keplerian
viscous accretion disk in vertical hydrostatic equilibrium.  We assume
that the disk is vertically isothermal, with a radial temperature
profile set by absorbing stellar radiation.  We further assume that
the disk is spatially thin ($z_H[r] \ll r$, where $z_H[r]$ is the
vertical pressure scale-height in the disk at a radial distance $r$
from the star), equivalent to neglecting all sources of pressure
support (e.g., thermal, magnetic) compared to rotation in the equation
of radial force balance.

With $\mstar$, $\lstar$, $r$ and the mean molecular mass $\mu$ of the disk gas normalised as: % (with the `twiddle' symbols denoting the normalized values):

$$ \tilde{\mstar} \equiv \frac{\mstar}{1 \msun} \,\,;\,\, \tilde{\lstar} \equiv \frac{\lstar}{1 \lsun} \,\,;\,\, \tilde{r} \equiv \frac{r}{1 {\rm AU}}\,\,;\,\, \tilde{\mu} \equiv \frac{\mu}{2.34}  \eqno(7) $$
the Keplerian angular rotation velocity becomes
$$ {\Omega}[\tilde{r}] = 2.0\times10^{-7} \left(\frac{\tilde{\mstar}}{{\tilde{r}}\,^3}\right)^{1/2} {\rm{s}}^{-1} \eqno(8)$$
and the radial temperature profile in the disk is
$$ {T}[\tilde{r}] = 393 \left(\frac{\tilde{\lstar}}{{\tilde{r}}\,^2}\right)^{1/4} {\rm K} \eqno(9)$$
where we have ignored any extra projection factors arising from the relative inclination between the disk surface and incident stellar radiation.  We note that \citet{1981PThPS..70...35H} (hereafter H81) invokes the same functional form for the temperature, but with a somewhat lower normalization of 280K based on the assumption that dust is severely depleted in the nebula while the gas absorbs only solar UV.  Many other recent works examining disk magnetic activity have adopted the latter standard Hayashi formulation \citep[e.g., B11,][]{2010ApJ...708..188T}.  Conversely, PC11 have employed the temperature structure of a passive flared disk, which has nearly the same functional dependence on radius ($T \propto r^{-3/7}$) but a much lower normalisation: $\sim$130K at 1AU around a solar mass star.  There are a number of uncertainties in determining what the disk temperature really is: for example, one most often assumes (as we have here) that the disk is passive, with negligible contribution from accretion heating, and is also vertically isothermal.  Without a rigorous reason for choosing a precise normalisation, we have opted here for the most naive estimate.  Since the reaction rates (\S4.2, Table 1) go as the square-root of the temperature, the difference between our normalization and e.g. that of H81 leads to negligible changes in the results, as we explicitly show in our comparisons to the results of other groups in \S6.1.  The sound speed is then
$$ c_s[\tilde{r}] = 6.1\times10^3 \left(\frac{T[\tilde{r}]}{\tilde{\mu}}\right)^{1/2} {\rm cm\,s^{-1}} = 1.2\times10^5 \left(\frac{\tilde{\lstar}}{{\tilde{\mu}}\,^4\,{\tilde{r}}\,^2}\right)^{1/8} {\rm cm\,s^{-1}} \eqno(10)$$
so that the disk pressure (and density) scale-height becomes
$$ z_h[\tilde{r}] \equiv \sqrt{2}\,\frac{c_s[\tilde{r}]}{\Omega[\tilde{r}]} = 8.5\times10^{11} \left(\frac{\tilde{\lstar}{\tilde{r}}\,^{10}}{{\tilde{\mu}}\,^4 {\tilde{\mstar}}^4}\right)^{1/8} {\rm cm} \eqno(11)$$
%Our normalization constant for the scale-height differs slightly from that of H81, due to the difference in our temperature normalizations.

Furthermore, the mass of a disk with a surface density profile $\Sigma [r]$ is given by:
$$M_d = \int_{R_{in}}^{R_{out}} (2\pi r\,\,\Sigma[r]\,\,{\rm d}r) \eqno(12)$$
We adopt a disk outer radius $R_{out} = {\rm constant} = 100 {\rm AU}$
over the range of stellar masses investigated here.  With insufficient
data at present about how this parameter changes with stellar mass,
this assumption, which sets $R_{out}$ to be of order that seen in
classical T Tauri stars, appears reasonable.  We also adopt a disk
inner radius $R_{in} \sim 5\rstar$, in agreement with magnetospheric
truncation models and observations; the precise value of this
parameter is immaterial to our calculations as long as $R_{in}$ lies
inside the radius where collisions of alkali metal atoms with other
species take over as the main ionization process, that is, where
thermal ionization dominates.  Finally, we assume that the disk mass
varies as $M_d \propto \mstar$.  This relationship, albeit with large
scatter (some of which may be age-related) is consistent with current
data \citep{2006ApJ...645.1498S, mohantysub}.  Specifically, we adopt
$M_d = 10^{-2} \mstar$, the median value indicated by observations of
solar-type classical T Tauri stars as well as young very low-mass
stars and brown dwarfs \citep{2005ApJ...631.1134A,
  2007ApJ...671.1800A, 2006ApJ...645.1498S, mohantysub}.

Our choices of $R_{in}$, $R_{out}$ and $M_d$ define the normalisation for any given surface density profile.  Here we examine a scaled version of the Minimum Mass Solar Nebula (hereafter MMSN; H81), with a radial surface density profile given by:
$$ \Sigma[\tilde{r}] = \Sigma_0 \,\tilde{r}\,^{-3/2} \,\,\,\,\, {\rm with} \,\,\,\,\, {\Sigma}_0 \equiv 7.1\times10^2 \tilde{\mstar} \,\,\,\,\,\,\,{\rm gm\,cm}^{-2} \eqno(13)$$
H81 gives a somewhat larger value of 1.7$\times$10$^3$ gm cm$^{-2}$ for the normalisation $\Sigma_0$ (for the Sun), because he uses essentially the same disk to stellar mass ratio as we do but a smaller outer disk radius ($R_{out}$ $\sim$ 36 AU instead of 100 AU).  Our volume density is then 
$$ \rho[\tilde{r},z] = {\rho}_0 \,\tilde{r}\,^{-11/4} \,{\rm e}^{-(z^2/z_h^2[\tilde{r}])} \eqno(14a)$$
where the scale-height $z_h[\tilde{r}]$ is given by equation (11), and 
$$ {\rho}_0 = 4.7\times10^{-10} \,\,{\tilde{\mstar}}\left(\frac{{\tilde{\mu}}\,^4{\tilde{\mstar}}^4}{\tilde{\lstar}}\right)^{1/8} {\rm gm\,cm}^{-3} \eqno(14b)$$
Note that our expressions for ${\rho}_0$ and $\Sigma_0$ include an extra factor of $\tilde{\mstar}$ compared to H81's formulation, due to the explicit dependence of our disk mass on the stellar mass.

\subsection{Ionization Rate and Chemistry}
We consider 3 sources of disk ionization: X-rays from the central star, cosmic rays, and the decay of radionuclides in the disk.  For the first, we estimate the total X-ray luminosity ($L_X$) from the empirical median relationship $L_X$/$\lstar$ $\sim$ $10^{-3.5}$ observed all the way from classical T Tauri stars to accreting brown dwarfs \citep[and references therein]{2007AA...468..353G}, and a suitable choice of $\lstar$ (described in \S5).  We consider coronal X-ray emission from an optically thin plasma dominated by impact excitation and ionisation by thermal electrons. Contributions from bound-bound, bound-free, and free-free radiation are included in the computation of our synthetic spectrum for all elements with atomic numbers 1 to 30, assuming the solar elemental abundances of \citet{1998SSRv...85..161G}. We use line and continuum emissivities from the CHIANTI compilation of atomic data \citep[and references therein]{2006ApJS..166..421L}, together with ion populations from Mazzotta et al.\,(1998), as implemented in the PINTofALE IDL software suite3 \citep{2000BASI...28..475K}. The spectra are calculated using an isothermal X-ray emitting gas of temperature ${\rm log}(T_X) = 7.2$, spherically distributed at a radius of 2$\rstar$ \footnote{This ignores the harder X-ray component observed in some cases, e.g., during flares; we intend to explore the additional effects of this hotter component in subsequent papers in preparation.}.  

X-ray ionization rates in the disc are subsequently computed with the {\sc moccasin} code \citep{2003MNRAS.340.1136E, 2005MNRAS.362.1038E, 2008ApJS..175..534E}, a 3-D photoionisation and dust radiative transfer code with a Monte Carlo approach to the propagation of ionising and non-ionising radiation. {\sc moccasin} has already been applied successfully to the modeling of X-ray irradiated protoplanetary discs \citep{2008ApJ...688..398E, 2009ApJ...699.1639E, 2010MNRAS.401.1415O}; more details regarding the physical processes and atomic data adopted for the calculations are given in \citet{2003MNRAS.340.1136E, 2008ApJS..175..534E} and \citet{2006MNRAS.372.1875E}. The first computations to address these issues were made by IG99, who also employed Monte Carlo methods to investigate the transport of X-rays within an idealised disk. Here we have repeated their calculations taking advantage of our newer code, the greater computing power available today, as well as a better understanding of the X-ray properties of YSOs granted by results from the Chandra and XMM-Newton space telescopes available over the last decade \citep{2005ApJS..160..353G, 2007AA...468..353G}. Moreover, {\sc mocassin} allows the inclusion of dust grains, the redistribution of secondary electrons and a full treatment of Compton scattering, and also uses a more sophisticated Monte Carlo estimator for the radiation field.  A detailed comparison of our rates to those of IG99 is the subject of a forthcoming paper (Ercolano et al 2012), where new parameterised curves for use in, e.g., MRI calculations, are also provided.  Here we simply note that for the {\it same (grain depleted) elemental abundances}, the ionisation curves we find and those computed by IG99 share the same physical characteristics, wherein a high ionisation rate is maintained in the upper tenuous disk regions before sharply falling off as the optical depth for absorption becomes of order unity, but the rate of decrease in our calculations is lessened by increased Compton scatterings due to the increased density.

Additionally, we follow \citet[hereafter UN09]{2009ApJ...690...69U} in calculating the cosmic ray ionization rate for molecular hydrogen as:
$$ \zeta^{H_2}_{CR}[r,z] = \frac{\zeta^{H_2}_{CR, ISM}}{2}\left[\rm{e}^{-\frac{\Sigma_a[r,z]}{\Sigma_{CR}}}\left\{ 1+\left(\frac{\Sigma_a[r,z]}{\Sigma_{CR}}\right)^{3/4}\right\}^{-4/3} \\
+ \rm{e}^{-\frac{\Sigma_b[r,z]}{\Sigma_{CR}}}\left\{ 1+\left(\frac{\Sigma_b[r,z]}{\Sigma_{CR}}\right)^{3/4}\right\}^{-4/3}\right] \eqno(15)$$
where $\Sigma_a[r,z] \equiv \int_z^\infty \rho[r,z]\,{\rm d}z$ is the surface density of the disk above the height $z$ at radius $r$, and $\Sigma_b[r,z] \equiv \int_{-\infty}^z \rho[r,z]\,{\rm d}z$ is similarly the surface density below the height $z$.  $\zeta^{H_2}_{CR, ISM}$ = 10$^{-17}$ s$^{-1}$ is the interstellar cosmic ray ionization rate for molecular hydrogen, and $\Sigma_{CR} \approx$ 96 g cm$^{-2}$ is the characteristic attenuation surface density for cosmic rays (UN09).  The two right-hand side terms in equation (15) thus express the total cosmic ray ionisation rate at a location within the disk as the sum of the rates due to attenuated cosmic rays arriving from the hemisphere above and the hemisphere below that location.  

Finally, we also follow UN09 in calculating the ionisation rate of molecular hydrogen due to the combined effect of short- and long-lived radionuclides as $ \zeta^{H_2}_{R} = 7.6 \times 10^{-19}$ s$^{-1}$.   

The total ionization rate $\zeta$ of H$_2$ is the sum of the X-ray, cosmic ray and radionuclide rates: $\zeta = \zeta^{H_2}_{X} + \zeta^{H_2}_{CR} +\zeta^{H_2}_R$.  For the ionization balance, we use a simplified chemical network, including both gas-phase and grain chemistry, essentially described by {\it model.4} of IN96.  The gas-phase chemistry in this model is that adopted by \citet[hereafter OD74]{1974ApJ...192...29O}, and schematically described by:
\bigskip

\begin{tabular}{|llclclclcl|}
\hline
\multicolumn{10}{|c|}{{\bf Table 1}: Gas Phase Chemistry (Reactions and corresponding Rate Coefficients)}\\
\hline
R1.  & m &  & & $\rightarrow$ & m$^{+}$ & + & e$^{-}$ & & $\zeta$ (ionization rate, s$^{-1}$)\\
R2.  & m$^{+}$ & + & e$^{-}$ & $\rightarrow$ & m & & & & $\tilde\alpha$ = $3 \times 10^{-6} / \sqrt{T}$ cm$^3$s$^{-1}$\\
R3.  & M$^{+}$ & + & e$^{-}$ & $\rightarrow$ & M & + & $h\nu$ & & $\tilde\gamma$ = $3 \times 10^{-11} / \sqrt{T}$ cm$^3$s$^{-1}$\\
R4.  & m$^{+}$ & + & M & $\rightarrow$ & m & + & M$^{+}$ & & $\tilde\beta$ = $3 \times 10^{-9}$ cm$^3$s$^{-1}$\\ 
\hline
\end{tabular}
\bigskip

\noindent where m and m$^+$ represent a molecule and its ionized counterpart, M and M$^+$ are a neutral heavy metal atom and its ionized counterpart, and e$^-$ is a free electron in the gas phase.  $\zeta$ is the total ionization rate for hydrogen molecules, as calculated above; the temperature for the reaction rates $\tilde\alpha$ and $\tilde\gamma$ is given by our radial temperature profile, equation (9).  

In our case (as in IN96), the metal is magnesium and the neutral molecule m is H$_2$.  However, the important {\it ionized} molecule m$^+$ is not H$_2^+$; while the ionization of H$_2$ initially creates this ion, the subsequent reaction chains invariably result in the dominant m$^+$ being an anion containing both hydrogen and some other heavier element(s), such as HCO$^+$, NH$^{4+}$ etc.  Following OD74, we only consider species involving hydrogen, carbon and oxygen for the molecular reactions, and omit nitrogen; we further omit those OD74 reactions involving pure atomic or molecular oxygen, following PC11 (see latter for justification).  In this case, the reactions R1 and R2 in Table 1 may be explicitly expanded into the following chain:
$$ {\rm H_2} + XCRR \rm \rightarrow H_2^+ + e^- \eqno{\rm R}(a) $$
$$ \rm H_2^+ + H_2 \rightarrow H_3^+ + H \eqno{\rm R}(b) $$
$$ \rm H_3^+ + CO \rightarrow HCO^+ + H_2 \eqno{\rm R}(c) $$
$$ \rm HCO^+ + e^- \rightarrow CO + H \eqno{\rm R}(d) $$
$$ \rm 2H + gr \rightarrow H_2 + gr \eqno{\rm R}(e) $$
where $XCRR$ signifies either an ionizing X-ray, cosmic ray or particle released by radionuclide decay, and gr represents a grain surface which catalyses the conversion of atomic to molecular hydrogen.  The cascade R($a$)--($c$) boils down, without any loss of generality, to:
$$ {\rm H_2} + XCRR \rm + CO \rightarrow HCO^+ + e^- + H \eqno{\rm R}(f) $$
which shows that the dominant molecular ion here is HCO$^+$.  Reactions R($f$), ($d$) and ($e$) capture the essence of our gas-phase chemistry: in the presence of ionizing influences and CO, molecular hydrogen is converted to HCO$^+$, electrons and atomic hydrogen; the HCO$^+$ and electrons dissociatively recombine to give back CO and more atomic hydrogen; and the circle is closed by the hydrogen atoms coalescing into hydrogen molecules on on the surface of grains that are well-mixed with the gas.  

However, this set presents a problem when the abundance of well-mixed grains becomes extremely low.  In such circumstances, the rate of conversion of H to H$_2$, reaction R($e$), can decrease to the point where atomic hydrogen dominates over molecular in the disk (which is indeed what happens in the dust-free simulations by \citet[hereafter BG09]{2009ApJ...701..737B}: X.\,Bai, pvt.\,comm.\,2011; see detailed discussion in \S6.1).  What is the threshold grain abundance below which this occurs?  We assume grains of uniform radius 0.1\,$\mu$m (see end of this section), and adopt the reaction rate formula supplied by BG09 (their equation (27)).  Then, for a fiducial disk temperature of 100\,K (achieved at $\sim$15\,AU around a solar-type star and 5\,AU around a very low-mass one), a fairly low gas number density of $n_H \sim 10^8$\,cm$^{-3}$ (attained at 2--3 scale heights at these radii), a sticking coefficient of $S_H\sim1$, and a constant H-to-H$_2$ conversion efficiency of $\epsilon\sim 10^{-3}$ (value preferred by BG09 after correcting for an error in the formulae by Cazaux \& Tielens (2004)), the timescale for converting initially completely atomic H entirely into H$_2$ reaches $\sim$10$^7$\,yr (the maximum lifetime of disks), for a dust-to-gas mass ratio of $R_{d:g}$ = 10$^{-5}$.  Thus, we nominally expect atomic hydrogen to dominate over H$_2$ in the upper/outer (low density) parts of a disk, when $R_{d:g} \lesssim 10$$^{-5}$. 

There are however a number of uncertainties in this calculation, most importantly in the conversion efficiency $\epsilon$.  On the one hand, Cazaux \& Tielens (2002, 2004) show that $\epsilon$ should increase rapidly with decreasing temperature (i.e., in the outer disk), compared to the fixed value adopted by us and BG09, thereby lifting the H$_2$ fraction.  On the other hand, we (in common with previous works such as BG09, B11 and PC11) have not considered the formation of ice mantles on grains beyond the ice line (i.e., at $T_{disk}$ $\lesssim 170$\,K).  H cannot be chemisorbed onto icy surfaces, only physisorbed; since such physisorption ceases at $\gtrsim$20\,K, H$_2$ formation on icy grain mantles halts at these temperatures (see discussion in Cazaux \& Tielens 2002).  Conversely, H$_2$ formation under such conditions may be mediated by other molecular species (Cazaux \& Tielens 2002, and references therein).       

Given these uncertainties, we {\it assume} that the gas always exists mainly as H$_2$, in accordance with the standard view of protoplanetary disks.  For the two higher dust abundance cases we examine, $R_{d:g} = 10^{-2}$ and 10$^{-3}$ (see end of this section), this assumption is consistent with the H-to-H$_2$ conversion rate discussed above.  For our two very low abundance cases, $R_{d:g} = 10^{-5}$ and 0, this amounts to assuming the presence of channels other than grain-surface recombinations for the conversion.  Imposing H$_2$ dominance by fiat, for all dust abundances, allows us to focus here on the primary process by which dust depletion affects the MRI: as grains are removed, the rate of electron adsorption onto grains (and the attendant rate of ion-electron recombinations on grains) nosedives, strongly altering the MRI efficiency.  If our calculations indicate that very low grain abundances are physically relevant (as turns out to be the case when we examine accretion rates; see point {\it (v)} in \S6.3), it will serve as an impetus to investigate the additional effect of hydrogen recombination channels at such abundances in more detail in future work.  We note that B11 similarly impose H$_2$ dominance in their grain-free case, for the same reason (discussed further in \S6.1).

To enforce H$_2$ dominance, we approximate the gas-phase reactions R($d$)--($f$) by the following set in all cases (both with and without grains):
$$ {\rm H_2} + XCRR \rm \rightarrow HCO^+ + e^- \eqno{\rm R}(g) $$
$$ \rm HCO^+ + e^- \rightarrow H_2 \eqno{\rm R}(h) $$
We also use R($h$) to approximate the dissociative recombination of HCO$^+$ and e$^-$ on grains, if the latter are present (we do account for the difference in recombination rates between the gas phase and grain surfaces).  Despite first appearances, this system preserves exactly the number of hydrogen particles: both the formation and dissociative recombination of an HCO$^+$ ion liberate an H each, which are simply lumped together as one hydrogen molecule in R($h$), expressing our condition that H$_2$ be the final state of any free hydrogen.  Reactions R($g$)--($h$) allow us to accurately follow the equilibrium abundance of the dominant molecular ion HCO$^+$, while enforcing the conversion of H into H$_2$ regardless of the grain abundance.  We do not follow the destruction and reformation of CO in this network, since HCO$^+$ is a trace species relative to CO.         

In summary, the gas-phase reactions R1 and R2 are interpreted as R($g$) and ($h$), with the stipulation that m = H$_2$ and m$^+$ = HCO$^+$; their rates remain as given in Table 1.   

As clear from the above, the chemical model further includes dust grains, and chemistry due to gas-grain as well as grain-grain collisions.  Grains are assumed to be either neutral, singly charged or doubly charged (henceforth gr0, gr$^{\pm}$ and gr$^{2\pm}$ respectively)\footnote{Ideally, higher grain charges should also be considered.  In reality, however, this is unlikely to make a significant difference to our results for 0.1\,$\mu$m sized grains: our derived electron abundances are in excellent agreement with those of PC11 for the same grain size (see discussion in \S6.1, point {\it (vi)}), even though the latter authors allow a much larger range of grain charges.}.  Reactions involving grains are divided into two classes: {\it mantle} chemistry, in which a neutral gas-phase particle is adsorbed onto a grain mantle (or the reverse process, desorption, in which an adsorbed neutral species is released into the gas phase); and {\it grain} chemistry, wherein a charged particle (whether gas-phase or grain) collides with a grain resulting in charge transfer.  We refer the reader to IN06 for the specific mantle and grain reactions considered (their Tables 3 and 4) and an exhaustive discussion of the model (reaction rates, sticking probabilities, desorption etc.).

Finally, the set of equations is closed by specifying the total number
density of particles in the gas-phase, $n$; the abundance of the heavy
metal species (Mg, both neutral and charged) relative to hydrogen,
$x[{\rm Mg}]$; the density and radius of each grain; and the
dust-to-gas ratio.  The chemical reactions cause negligible variations
in the mean molecular weight $\mu$ of the gas (IN06), so we adopt a
constant $\mu$ = 2.34, corresponding to a solar composition for the
most abundant elements, H and He.  The number density of gas-phase
particles at any location is then $n[r,z] \approx \rho[r,z]/(\mu
m_H)$, with $\rho$ given by equation (14) and $m_H$ the mass of a
hydrogen nucleus.  For the gas-phase abundance of Mg we adopt $x$[Mg]
= $6.78 \times 10^{-9}$ per H$_2$ particle (10$^{-4}$ solar).

Real disks have a distribution of grain sizes, determined by the balance between collisional coagulation and fragmentation.  Recent simulations by Birnstiel et al. (2010) indicate that fragmentation results in a significant population of small grains even in the face of strong grain growth.  Moroever, while larger grains quickly settle to the disk midplane, very little turbulence is needed to keep the smallest ones aloft and well-mixed with the gas.  Overall, therefore, one expects the bulk of the dust {\it mass}, locked in larger grains, to be rapidly confined close to the midplane, while a significant fraction of the grains in terms of {\it number}, skewed towards the smallest sizes, remains mixed throughout the disk. Finally, the smallest grains are the most important in the grain reactions discussed above: the latter depend crucially on the grain surface area per unit gas mass, which is largest for the smallest grains.  

We encapsulate this picture as simply as possible in our calculations, as follows.  We consider only very small ISM-sized grains, of uniform radius $a$ = 0.1 $\mu$m and internal density $\rho_{gr}$ = 3 g cm$^{-3}$, well-mixed with the gas.  At the same time, we vary the dust-to-gas mass ratio from run to run, using the values 1:100 (standard ISM), 1:$10^3$ (90\% depleted relative to standard ISM), 1:$10^5$ (99.9\% depleted) and 0 (100\% depleted).  The ``depleted'' cases simulate conditions wherein the dust mass is increasingly bound up in larger grains that have settled to the midplane and thus no longer interact with the gas.  The 100\% depletion case is not expected to hold literally in real disks, but rather represents the limiting value.  A more realistic treatment demands dynamic calculations that account for the time-dependent effects of grain growth and settling \citep{2010ApJ...708..188T}; this will the subject of future work.     

The methods we use to find chemical equilibrium are entirely standard.  We time-integrate the chemical kinetic equations by semi-explicit extrapolation, an algorithm that works well with such stiff ODEs and is described in Numerical Recipes \citep{1992nrfa.book.....P}.  The integration is stopped when all species' timescales for further change become much longer than the disk lifetime.  Substituting the final abundances back into the kinetic equations in a few specific cases demonstrates that the solutions are very close indeed to equilibrium.

\subsection{Resistivities}
Armed with the equilibrium abundances of the various species from the chemical network calculations, we compute the resistivity at every location, and thereby investigate where the disk is MRI active by the criteria of \S2 for a given field strength $B$.  We follow \citet{2007ApSS.311...35W} in writing the Ohmic, Hall and Pedersen conductivities ($\sigma_O$, $\sigma_H$ and $\sigma_P$ respectively) as
$$ \sigma_O = \frac{ec}{B}\sum\limits_j n_j |Z_j| \beta_j \eqno(16) $$
$$ \sigma_H = \frac{ec}{B}\sum\limits_j \frac{n_j Z_j}{1+\beta_j^2} \eqno(17) $$
$$ \sigma_P = \frac{ec}{B}\sum\limits_j \frac{n_j |Z_j| \beta_j}{1+\beta_j^2} \eqno(18) $$
where the summation is over all charged species $j$ with particle mass $m_j$, number density $n_j$ and charge $Z_j e$. The Hall parameter $\beta_j$ (not to be confused with the plasma $\beta$ parameter) is the ratio of the gyrofrequency of a particle of species $j$ to its collision frequency with neutrals (of mean particle mass $m_n = \mu m_H$ and density $\rho_n$):
$$ \beta_j = \frac{|Z_j|eB}{m_j c}\frac{1}{\gamma_j \rho_n} \eqno(19) $$
Here $\gamma_j = <\sigma v>_j/(m_j + m_n)$ is the drag coefficient and $<\sigma v>_j$ the rate coefficient for collisional momentum transfer between species $j$ and the neutrals, making $\gamma_j \rho_n$ the collision frequency with neutrals.  This allows the induction equation to be written in the form:
$$ \frac{\partial {\mathbf {B}}}{\partial t} = {\mathbf \nabla}\times({\mathbf v}\times{\mathbf B}) - {\mathbf \nabla}\times[\eta_O ({\mathbf \nabla}\times{\mathbf B}) + \eta_H({\mathbf \nabla}\times{\mathbf B})\times\hat{\mathbf B} + \eta_A({\mathbf \nabla}\times{\mathbf B})_{\perp}] \eqno(20) $$  
where `` $\hat{}$ '' denotes a unit vector and `` $\perp$ '' indicates the component of a vector perpendicular to $\mathbf{B}$.  The first expression on the right is the inductive term ($I$), while the second, third and fourth are the Ohmic ($O$), Hall ($H$) and ambipolar ($A$) resistive terms respectively.  The ratio of the inductive term to whichever of the resistive terms is dominant is thus of the form $Lv/\eta \sim L^2/t\eta$, where $L$ is the field diffusion lengthscale and $t$ the field advection (growth) timescale, and the resistivity $\eta$ equals $\eta_O$, $\eta_H$ or $\eta_A$; for the field to grow, one expects this ratio to exceed unity.  $L \sim v_K/\Omega$ for large-scale fields generated by shearing radial ones, while $L \sim v_{\mathcal A}/\Omega$ for local fields created by the MRI; in both cases $t \sim 1/\Omega$.  This yields equations (1) and (2) when Ohmic diffusion holds sway ($\eta = \eta_O$), and the naive condition equation (3) when ambipolar diffusion rules instead ($\eta = \eta_A$).   The resistivities are:
$$ \eta_O = \frac{c^2}{4\pi\sigma_O} \eqno(21) $$
$$ \eta_H = \frac{c^2}{4\pi\sigma_\perp}\frac{\sigma_H}{\sigma_\perp} \eqno(22) $$
$$ \eta_A = \frac{c^2}{4\pi\sigma_\perp}\frac{\sigma_P}{\sigma_\perp} - \eta_O \eqno(23) $$
with $\sigma_\perp \equiv \sqrt{\sigma_H^2 + \sigma_P^2}$ being the total conductivity perpendicular to the magnetic field.  Note, from equations (16)--(19) and (21)--(23), that: {\it (i)} $\eta_O$ is independent of the field strength $B$, while $\eta_H$ and $\eta_A$ scale linearly and quadratically, respectively, with $B$; and {\it (ii)} {\it if} ions and electrons are the the only charged species ({\it not} true in our case), then equation [25] reduces to $\eta_A = \beta_i \beta_e \eta_O$, yielding (using $\beta_i \ll \beta_e$) the simplification $v_{{\mathcal A}z}^2/\eta_A\Omega \approx \gamma_i\rho_i/\Omega$, as advertised in \S1.2. To compute the resistivities, we must specify the rate coefficients $<\sigma v>_j$ for $j$ = $e$ (electrons), $i$ (ions) and $gr$ (grains).  We employ the values supplied by \citet{1999MNRAS.303..239W}:
$$ <\sigma v>_e \,= 10^{-15} \left(\frac{128kT_e}{9\pi m_e}\right)^{1/2} \,\,\,\,{\rm cm}^3 \,{\rm s}^{-1} \eqno(24) $$
$$ <\sigma v>_i \,= 1.6 \times 10^{-9} \,\,\,\,{\rm cm}^3 \,{\rm s}^{-1} \eqno(25) $$
$$ <\sigma v>_{gr} \,= \pi a^2 \left(\frac{128kT_n}{9\pi m_n}\right)^{1/2} \,\,\,\,{\rm cm}^3 \,{\rm s}^{-1} \eqno(26) $$
where $T_e$ and $T_n$ are the electron and neutral temperatures respectively, and $a$ is the grain radius (fixed at 0.1 $\mu$m in our case).  Note that $<\sigma v>_i$ is the same for molecular and metal ions.  We assume that $T_e$ = $T_n$ = $T[r]$, with the latter given by equation (9).

\section{Stellar Models Considered}
We examine two stellar models.  The first corresponds to a typical classical T Tauri star: $\mstar$ = 0.7$\msun$ , $\rstar$ = 2$\rsun$ and $T_{\ast}$ = 4000K.  The second corresponds to a very low mass star, also in its disk accretion phase:  $\mstar$ = 0.1$\msun$ , $\rstar$ = 1$\rsun$ and $T_{\ast}$ = 3000K.  The radii and temperatures adopted for both masses are from the Lyon evolutionary tracks \citep{1998AA...337..403B} for an age of 1 Myr, appropriate for objects in their classical T Tauri phase.

These numbers correspond to bolometric luminosities of $\lstar$ = 0.92 and 0.07 $\lsun$ for $\mstar$ = 0.7 and 0.1 $\msun$ respectively.  Our adopted scaling $L_X/\lstar = 10^{-3.5}$ then yields an X-ray luminosity of $L_X$ = $1.12 \times 10^{30}$ erg s$^{-1}$ for 0.7$\msun$, and $L_X$ = $8.84 \times 10^{28}$ erg s$^{-1}$ for 0.1$\msun$.    

\section{Results}

\subsection{Ionization Balance and Chemistry}
The equilibrium fractional abundances of all the species we have considered, defined as $x_a \equiv n_a/n_{{\rm H}_2}$ (where $n_a$ and $n_{{\rm H}_2}$ are the number densities of species $a$ and H$_2$ molecules respectively), are plotted in Figs.\,1 and 2.  Fig.\,1 shows the situation at a radial distance of 1\,AU, for stellar masses of 0.7\,$\msun$ (top panels) and 0.1\,$\msun$ (bottom panels), and dust-to-gas ratios (assuming 0.1\,$\mu$m grains) ranging from 10$^{-2}$ (standard ISM) to zero (left to right).  Fig.\,2 shows the same for 10\,AU.  Some clear trends are noteworthy, as follows.    

\noindent {\it (i)} At any radius, the inclusion of even a tiny fraction of well-mixed 0.1\,$\mu$m grains drastically suppresses the abundance of free electrons as one moves towards the midplane, relative to the no-dust case.  Moreover, the height at which the drop-off commences and the steepness of the decline both increase with grain abundance.  In other words, in the presence of grains, $x_e$ decreases dramatically with both increasing density and increasing grain abundance.  

\noindent {\it (ii)} In low density regions of the disk, i.e., sufficiently high above the midplane, charge equilibrium is dominated by the balance between free electrons and molecular ions (HCO$^+$ in our case), independent of the grain abundance. 

\noindent {\it (iii)} The abundance of metal ions, $x_{Mg^+}$, falls off precipitously towards the midplane with increasing grain abundance.

\noindent {\it (iv)} Related to the above, we see that in the absence of grains, charge equilibrium close to the midplane (i.e., at high densities) is maintained by the balance between metal ions and free electrons.  At sufficiently high grain abundances, on the other hand, metal ions become negligible, and charge equilibrium near the midplane is maintained instead by either the balance between charged grains, or between free electrons and molecular ions (HCO$^+$).  

BG09 and \cite{2009ApJ...698.1122O} have previously noted the above
trends for solar-type stars; our analysis reproduces their results for
such stars ($\sim$0.7\,$\msun$), and shows that these trends apply to
disks around very low-mass stars ($\sim$0.1\,$\msun$) as well.

\noindent {\it (v)} The reason for the decrease in metal ions with increasing grain abundance has been debated before. IN06 argue that the decline is due to the sweeping up (via adsorption) of Mg atoms by grains, preventing the formation of Mg$^+$.  In contrast, BG09 posit that it is because grains promote the recombination of metal ions into neutral atoms.  Our analysis shows that both effects can be at play, depending on the disk conditions, as follows.

To begin with, note that at 1\,AU for 0.7\,$\msun$, the abundance of free (gas-phase) neutral Mg atoms close to the midplane is hardly affected by the grain abundance, while the abundance of Mg adsorbed onto grains (i.e., grMg) is orders of magnitude below that of the free neutral Mg.  Clearly, most of the neutral Mg is not swept up by grains, and thus such adsorption does not {\it cause} the steep decline in Mg$^+$.  However, the abundance of adsorbed Mg does increase as the Mg$^+$ and electron abundances decrease.  The only explanation is that grains facilitate the recombination of metal ions into neutral atoms, which are then adsorbed onto the grains.  In other words, the cause of the decline in Mg$^+$ here is recombination stimulated by grains, and the increase in adsorbed Mg is a {\it result} of this process.  This is consistent with the scenario proposed  by BG09.  The same is true at 1\,AU for 0.1\,$\msun$ as well (the abundance of adsorbed Mg is much higher than in the 0.7\,$\msun$ case, but still far below that of neutral Mg, which remains unaffected near the midplane by the grain abundance).    

At 10\,AU, the situation is radically different.  As the grain abundance increases, the fractional abundance of adsorbed Mg becomes comparable to, and eventually much higher than, the abundance of gas-phase neutral Mg (at dust-to-gas ratios of $\gtrsim 10^{-3}$ for 0.7\,$\msun$, and $\gtrsim 10^{-5}$ for 0.1\,$\msun$).  At the lower temperatures prevalent at this radius, grains do sweep up much of the neutral Mg, directly preventing the formation of Mg$^+$, as IN06 propose.  This effect is also much stronger for the 0.1\,$\msun$ case compared to 0.7\,$\msun$.      

\noindent {\it (vi)} 	Our quantitative electron abundances in the grain-free case agree with the results of PC11 and BG09, but differ significantly from those of B11, for a subtle reason.  

To illustrate this, we first plot $x_e$ and the ionization rate $\zeta$ as a function of the total number density of particles, at 1\,AU for a 0.7\,$\msun$ star, for both the grain-free case and a standard dust-to-gas ratio of $10^{-2}$ (Fig.\,3a).  For comparison, we refer to Fig.\,12 by PC11.  In the latter, PC11 plot $x_e$ as a function of the number density of H$_2$ (essentially equal to the total number density), for a fixed disk temperature of 280\,K (corresponding to 1\,AU for a solar-type star in the H81 model) and a constant $\zeta = 10^{-17}$\,s$^{-1}$.  They show the results for both their own chemical network as well as the simple and complex networks of BG09.  

Our Fig.\,3a shows that $\zeta = 10^{-17}$\,s$^{-1}$ corresponds to $n  \approx 2\times10^{13}$\,cm$^{-3}$ in our model at 1\,AU.  For this density, we find $x_e \approx 10^{-15}$ for a gas-to-dust ratio of $10^{-2}$ with 0.1\,$\mu$m grains, and $x_e \approx 3\times10^{-10}$ without grains.  This is in excellent agreement with the $x_e$ plotted by PC11 for the same density and same grain conditions.  Specifically, our $x_e$ almost exactly equals the PC11 network result at this density for both the dust-free and dusty cases; it also equals the BG09 results for the dust-free case, and is within a factor of $\sim$3 of the latter for the standard dust-to-gas ratio.  The agreement with PC11 is unsurprising, given that our chemical networks are nearly identical.  Importantly, this also shows the negligible impact of the difference in disk temperature between our models (393\,K versus 280\,K at 1\,AU).   

Next, we compare to the results of B11.  To make this straightforward, we replot our $x_e$ and $\zeta$ at 1\,AU for the 0.7\,$\msun$ case as a function of fractional scale height ($z/z_H$) in Fig.\,3b, and compare to the analogous plot in Fig.\,1 by B11.  Our ionization rates due to cosmic rays and radionuclides differ somewhat from B11's, so it is more instructive to compare our results in the regions where X-ray ionization dominates, i.e., $\zeta \gtrsim 10^{-17}$\,s$^{-1}$.  For specificity, consider again the exact value $\zeta = 10^{-17}$\,s$^{-1}$, reached in both our case and B11's at $z/z_H \sim 2$ (marked in our plot as well as B11's by a vertical dashed line).  For the case of 0.1\,$\mu$m grains with a dust-to-gas mass ratio of $10^{-2}$, B11 finds $x_e \approx 3\times10^{-16}$ at this $\zeta$, roughly comparable to our value of $10^{-15}$.  For the dust-free case, however, he finds $x_e \approx 10^{-12}$, more than 2 orders of magnitude below our value of $3\times10^{-10}$, and thus {\it also} similarly lower than the PC11 and BG09 predictions.  The difference appears to be due to a subtle change in chemistry in the grain-free case between the earlier analysis by BG09 and the new one by B11 (the following discussion was greatly aided by X.\,Bai, private comm., 2011).  

In particular, BG09 and B11 differ in their treatment of H$_2$ formation.  Some of the intermediate chemical reactions lead to the formation of atomic hydrogen, requiring a channel to convert this back to H$_2$ (see discussion in \S4.2).   BG09 assume that the recombination of atomic to molecular hydrogen occurs on grain surfaces.  In the {\it absence} of grains, however, most of their hydrogen remains in atomic form.  B11, aware of this issue, artificially adds a reaction to convert H to H$_2$ independent of grains (X.\,Bai, pvt.\,comm.\,2011), to ensure that H$_2$ is the dominant form in all cases.  This in turn increases the NH$_4^+$ abundance in his chemical network by orders of magnitude compared to BG09, and thereby decreases the free electron fraction by a comparable amount due to the dissociative recombination of NH$_4^+$.  This explains the difference between BG09 and B11.    

Conversely, our calculations (and those of PC11) have only HCO$^+$ and no NH$_4^+$, but we also enforce by {\it fiat} (essentially as B11 does) that any atomic hydrogen is converted to H$_2$ even for the grain-free case (see \S4.2).  We therefore see that the combination of mainly atomic hydrogen and the inclusion of NH$_4^+$ serendipitously leads BG09 to approximately the same $x_e$ as we (and PC11) get with mainly molecular hydrogen in the absence of NH$_4^+$, but the combination of molecular hydrogen with NH$_4^+$ leads B11 to much lower $x_e$.  A full resolution of this issue requires a better understanding of hydrogen recombination channels for extremely low grain abundances.  

Nevertheless, as noted earlier, even a small abundance of well-mixed grains strongly affects the electron fraction.  In the presence of dust, as shown, our results are consistent with those of B11 as well as BG09 and PC11, despite variations in the chemistry.

\subsection{MRI Active Zones}
Next, we calculate the location and vertical extent of the MRI-active
zone at every radius, using the stress-maximizing technique described
in \S3.  Before discussing our full disk results, we illustrate the
technique for the fiducial case of 0.7\,$\msun$ with a dust-to-gas
ratio of $10^{-2}$, at disk radii of 1\,AU and 10\,AU (Figs.4a and b
respectively).

\subsubsection{Illustrative Results at 1 and 10\,AU for 0.7\,$\msun$ and Standard Dust Abundance}

Consider the 1\,AU case first.  The top left panel of Fig.\,4a shows $hB^2$, where $h$ is the thickness of the active layer, as a function of the field strength $B$.  We see that this quantity increases monotonically with $B$, and then drops extremely rapidly to zero beyond its maximum value, as $\beta$ falls below $\beta_{\rm min}$ for stronger fields and the MRI is quenched at all vertical locations at this radius.  In other words, the maximum value of $hB^2$ occurs essentially at the $B_{\rm max}$ allowed for a given radius; hence, for all practical purposes, we are justified in setting $B_{\rm max}$ (which corresponds to $\beta_{min}$) precisely equal to the (very slightly lower) field strength at which the height-integrated stress $hB^2$ is maximized, as stated in \S3.  

The top right panel of Fig.\,4a explicitly shows the vertical location of the MRI-active region as a function of the field strength, and the conditions that constrain its thickness.  The horizontal dashed line marks the particular value $B_{\rm max}$, at which $hB^2$ is maximized in the previous panel.  We see that the upper envelope of the active layer is always determined by the limiting ambipolar condition $\beta = \beta_{\rm min}$.  Conversely, for small $B$, the lower envelope is set by the minimum Ohmic Elsasser number criterion, $\Lambda_{\mathcal A} = 1$.  As the field strength increases to nearly $B_{\rm max}$, however, the lower envelope of the active region instead becomes determined by the ambipolar criterion $\beta = \beta_{\rm min}$ as well.  This is not surprising: given that for fields stronger than $B_{\rm max}$, the ambipolar condition shuts off MRI {\it everywhere} (by the definition of $B_{\rm max}$), it is to be expected that, close to $B_{\rm max}$, this criterion supersedes the Ohmic one even close to the midplane.  This is ultimately because the ambipolar resistivity $\eta_A$ increases quadratically with $B$, while $\eta_O$ is independent of $B$.  While B11 does not explicitly mention this effect, it is present in his grain-inclusive calculations as well, as evinced by the characteristic ``dome-shaped'' peak of the active region close to $B_{\rm max}$ in both his plots and ours (see his Fig.\,3, for 0.1\,$\mu$m grains at 1 and 10\,AU), arising from the ambipolar condition sculpting both the upper and lower envelopes of this region.  Finally, the vertical line shows the height at which the magnetic Reynolds number criterion for large-scale fields achieves its limiting value, $\Lambda_K = 10$.  Note that this location is independent of the field strength.  Moreover, for any value of $B$ for which an active layer exists, we see that the Elsasser number criterion is achieved higher in the disk (i.e., at smaller $\eta_O$) than the Reynolds condition, which reflects the fact that $v_K \gg v_{\mathcal A}$ for a thin disk in which $\beta \leq \beta_{\rm min} < 1$ (see discussion in \S2).  

The bottom left panel of Fig.\,4a shows the values of the MRI-activity criteria, $\Lambda_K$, $\Lambda_{\mathcal A}$ and $\beta/\beta_{\rm min}$, as a function of height for the optimal field strength $B_{\rm max}$.  The active region ($\Lambda_K > 10$, $\Lambda_{\mathcal A} > 1$ and $\beta/\beta_{\rm min} > 1$) is sandwiched between $z/z_H \sim$ 2.5 and 3.5 (the intersection of the $B = B_{\rm max}$ line and the active zone in the previous panel).  As noted above, we also explicitly see that both the upper and lower limits of this layer are set by the ambipolar criterion: the Ohmic Elsasser number exceeds unity slightly deeper in the disk than the actual bottom of the active region.  Finally, the far less stringent Reynolds criterion is satisfied much deeper in the disk.  Regions closer to the midplane than the locations of both the Reynolds and Elsasser limiting values comprise the Dead Zone, regions that lie between the two make up the Undead Zone, and regions where the ambipolar criterion is not satisfied (everywhere outside the active layer in this case) constitute the Zombie Zone.          

The bottom right panel of Fig.\,4a shows the Ohmic, Hall and ambipolar
resistivities as a function of height, for $B_{\rm max}$.  We see that
the curves for $\Lambda_{\mathcal A}$ and $\beta/\beta_{\rm min}$, in
the previous panel, essentially reflect the inverse of $\eta_O$ and
$\eta_A$, as expected.  $\eta_O$, which depends on the ionization
fraction, decreases with increasing height (as the electron abundance
increases), while $\eta_A$, which depends on both the ionization
fraction and the density, first decreases with increasing height (as
$x_e$ increases) and then rises again (as the density continues to
decrease).  Note that $\eta_H$ changes sign at $z/z_H \sim 5$.  Note
also the flattening of all the resistivities below $z/z_H \sim 2$,
which is caused by a flattening of the ionization rate close to the
midplane (where the ionization is mainly due to cosmic rays and
radionuclides, at this radius; e.g., see Fig.\,3b).  The slight jitter
in $\eta_A$ very close to the midplane is due to small number photon
statistics, at these very high optical depths, in our Monte Carlo
transfer simulations.

The results for 10\,AU, plotted in Fig.\,4b, are qualitatively very similar, albeit with three differences worth mentioning.  First, the range of field strengths for which an MRI-active layer exists, and thus $B_{\rm max}$, is significantly smaller than at 1\,AU.  This is because the gas pressure $P_{gas}$ at all heights at this larger radius is lower, and thus the magnetic pressure $P_B$ for which $\beta \geq \beta_{\rm min}$ is also smaller.  Second, $\Lambda_K > 10$ at all heights at 10\,AU (which is why the curve for this criterion does not appear in the second and third panels of Fig.\,4b), meaning that there is no Dead Zone at this radius, only Undead and Zombie Zones.  Third, there is no flattening of the $\eta$ curves close to the midplane, unlike in the 1\,AU case: the lower densities at 10\,AU allow X-rays to penetrate much more easily to the midplane, and thus dominate over cosmic rays and radionuclides in controlling the ionization rate there.      

These results are very similar to those of B11 for the same stellar type, disk radii and grain properties.  The only small difference is that our $B_{\rm max}$ are somewhat greater than his (by 3--5$\times$) at a given radius, because of differences in chemistry, and also because our ionization rates are somewhat higher for the same ionizing flux (due to a slightly lower surface density in our case compared to the standard MMSN -- see discussion of equation (13) in \S4.1 -- and also due to our inclusion of Compton scattering in a self-consistent manner, which makes our X-ray ionization rate fall off slower with increasing optical depth compared to the results of IG99, which B11 uses for his analysis).  Conversely, as discussed in \S6.1, our electron abundances close to the midplane for a {\it dust-free} case are considerably higher than B11 finds; for this specific idealized situation, our $B_{max}$ are correspondingly significantly higher (not plotted), because both $\eta_O$ and $\eta_A$ are smaller when more free electrons are present.  Finally, our results for the 0.1\,$\msun$ case are qualitatively similar to, with the same trends as, the ones plotted in Fig.\,4 for 0.7\,$\msun$.   

\subsubsection{Full Disk Results} 

With these preliminaries dealt with, we move on to a discussion of our full disk results.  In Figs.\,5--8 we plot our results for $\mstar = 0.7$\,$\msun$, for dust-to-gas ratios of $10^{-2}$--0; Figs.\,9--12 show the same for 0.1\,$\msun$.  The field strength adopted at each radius is the optimal $B_{\rm max}$ derived as discussed above.   The top panel in each plot illustrates the relative importance of the various resistivities as a function of radius and column depth, the middle panel additionally shows the locations of the Dead and Undead Zones, and the bottom panel shows the location of the Zombie Zone as well.  The final MRI-active layer in each case is the region bounded by black filled circles in the bottom panel.  We note that there are a few sharp discontinuities/jitters in the boundary of the active layer at large radii in some cases, marking a transition in the shape of the layer with radius.  The changes in the shape are themselves real; they reflect the fact that the resistivities vary with radius and height, sometimes very rapidly (as illustrated in Figs.\,1 and 2).  The few discontinuities and sharp fluctuations at these transitions, however, are not real, but rather due to the inherent statistical uncertainty in our Monte Carlo determination of the ionization rate.  Fewer photon packets are incident on the disk at large radii, yielding a relatively higher jitter in the derived ionization rates there.  The resulting small statistical oscillations in the inferred resistivities can lead to somewhat larger fluctuations in the active layer boundary, by pushing the diffusivity over the critical threshold that defines the boundary.  We have made no attempt to remove these jitters, since they are a measure of the accuracy of our results (and smoothing them out by using more X-ray packets would have entailed prohibitively long processing times).  Nevertheless, the overall shape of the active layer as a function of radius is clear in all cases, and the similarity in shapes between the 0.7 and 0.1\,$\msun$ simulations for a fixed dust-to-gas ratio (and the agreement with the semi-analytic results of B11 for the dusty solar-type case, discussed above) implies that our results are physical, and not dictated by statistical errors.  

There are five main conclusions that we may draw from these plots.  

\noindent {\it (i)} For both 0.7 and 0.1\,$\msun$, the Dead and Undead Zones become smaller with decreasing grain abundance, and essentially disappear in the dust-free case.  This is a consequence of increasing $x_e$, and thus decreasing Ohmic resistivity, with declining amounts of dust.   

\noindent {\it (ii)} For a fixed dust-to-gas ratio, the radial extents of the Dead and Undead Zones are smaller for 0.1\,$\msun$ than for 0.7\,$\msun$, by a modest factor of $\sim$1.5--2.  The X-ray ionizing flux from the very low-mass star is far lower than that from the solar-type one, but the gas density (and hence X-ray optical depth) at any radius is also much lower in the disk around the former; the two effects compensate for each other in setting the ionization rate, making the Dead and Undead Zones roughly comparable in size between the two stars.  For the solar-type case, our Dead and Undead Zone extents (e.g., $\sim$3 and 13\,AU respectively for a dust-to-gas ratio of $10^{-2}$) are also consistent with those found by \citet[who use somewhat larger 1\,$\mu$m grains, and a magnetic pressure fixed at 0.1\% of the midplane gas pressure]{2009ApJ...703.2152T}.        

\noindent {\it (iii)}  In all cases, both the upper and lower boundaries of the MRI-active layer are set by the minimum ambipolar condition $\beta = \beta_{\rm min}$.  At radii where the Undead Zone exists, its upper envelope (set by the Elsasser number limit $\Lambda_{\mathcal A} = 1$) roughly follows, but still lies below, the lower envelope of the active layer.  This has an important consequence for mass accretion.  The raison d'\^{e}tre for the Undead Zone is that the MRI is not required for angular momentum transport within this zone: the transport here occurs via large-scale toroidal fields generated from seed radial ones by Keplerian shear, {\it as long as the Undead Zone is able to siphon off seed fields from neighboring MRI-active areas}.  If this zone is separated from the MRI-active region by an ambipolar-dominated {\it inactive} layer, however, such siphoning may be quenched.  In this case, the Undead Zone effectively dies as well, and cannot be invoked to supplement the MRI-driven angular momentum transport and mass accretion.                     

\noindent {\it (iv)}  For a fixed dust-to-gas ratio, the shape and location of the active layer is roughly similar for 0.1 and 0.7\,$\msun$.  The midplane in the outer disk is MRI active for all the dust abundances considered, while activity extends to the midplane in the inner disk only in the dust-free case.  For both stars, the active layer moves closer to the midplane with decreasing dust abundance (because the fraction of free electrons increases at all heights, and thus $\eta_A$ decreases); simultaneously, it also becomes thinner, in units of decades of column density.  

\noindent {\it (v)}  The combined effect of the two trends in {\it (iv)} is that the total mass of the active layer remains roughly constant with varying dust-to-gas ratio, comprising $\sim$5--10\% of the total disk mass for both 0.1 and 0.7\,$\msun$.             

\subsection{Accretion Rates}
Having determined the locus of the active layer in each disk, we estimate the mass accretion rate through it.  As discussed in \S3, B11 adopts $\mdot = hB_{\rm max}^2/4\,\Omega$, which only holds for a radially constant rate, while we use the general expression $\mdot = (1/2\,r\Omega)\,\partial(r^2 h B_{\rm max}^2)/\partial{r}$.  Figs.\,13--16 show our results for dust-to-gas ratios of $10^{-2}$--0.  In each plot, the top panels illustrate various relevant quantities as a function of disk radius for 0.7\,$\msun$, while the bottom panels show the same for 0.1\,$\msun$.  For each star, we plot from left to right: {\it (1)} $B_{\rm max}$; {\it (2)} $hB_{\rm max}^2$, which determines the accretion rate (modulo $\Omega$) in B11's case; {\it (3)} $r^2hB_{\rm max}^2$, the {\it slope} of which determines the accretion rate (modulo $r\Omega$) in our case; and {\it (4)} the $\mdot$ derived from these quantities, using both B11's formula and ours.  As an aside, notice that the shape of the B11 $\mdot$ as a function of radius is similar to that of $r^2hB_{\rm max}^2$; this merely reflects the fact that the former is proportional to $r^{3/2}hB_{\rm max}^2$, only a factor of $r^{1/2}$ different from the latter.  We further note that we have evaluated the differential in our $\mdot$ expression via simple finite differencing over our radial grid, which leads to small-scale unevenness in the derived accretion rate.  We have not attempted to smooth this out, since the overall trends in our $\mdot$ -- average magnitude as a function of radius, steep rise/falloff at certain locations, and a reversal in the direction of the flow at others, all of which {\it are} real, as discussed below -- remain clear.  There are six main implications of these plots.  

\noindent {\it (i)}  The $\mdot$ inferred with B11's formula are radially variable.  While B11 discusses some physical consequences of this, the bare fact of radial variations demonstrates the inconsistency in using this formula, which is only valid for $\mdot$ constant with $r$.  

\noindent {\it (ii)}  Our $\mdot$, derived from the general equation, is also radially variable; moreover, it also changes {\it sign} at some locations.  That this must happen is easily verified by eye from the plot of $r^2hB_{\rm max}^2$ in the third panel: where this quantity increases with radius, the accretion rate, proportional to the slope of this function, is positive; where it is flat, the rate drops to zero; and where it decreases with radius, the rate is negative, i.e., material flows outwards.  

\noindent {\it (iii)}  The immediate consequence of these radial variations in $\mdot$ is that material builds up in some regions of the active layer, and is preferentially evacuated from others.  Whether this process runs away, oscillates or reaches a steady state, however, is complicated, as follows.  

Consider, for example, a disk radius at which the active layer does not extend all the way to the midplane, and where moreover there is a net influx of matter ($\mdot$ coming into the region exceeds that going out). The latter enhances the density at every vertical location inside the active layer here.  Therefore, at any given time, and for a fixed field strength, the new ionization and chemical equilibrium at each height within the layer here equals that at a lower (denser) height at an earlier time\footnote{Note that a quasi-steady ionization and chemical equilibrium is always expected, because the recombination and field growth/diffusion timescales are much shorter than the secular timescales over which mass accretion occurs; see B11.}.  But the bottom of the active layer is originally set by the density above which (or more precisely, the ionization fraction below which) the MRI cannot be sustained for $B_{\rm max}$.  Thus, with increasing density (and hence decreasing $x_e$), the bottom of the layer moves upwards, while the location of the top remains fixed for the given field strength (since there is no change in density at any point originally above the active layer).  Nor can the field strength change from the initial $B_{\rm max}$: it cannot get weaker, because we have assumed that the MRI maintains the strongest possible field that still allows the MRI to operate, and it cannot get stronger, because $B_{\rm max}$ is already by definition the strongest (vertically constant) field that can be maintained here.  The upshot is that the active layer keeps getting thinner, and thus the quantity $hB_{\rm max}^2$ continues to diminish, at this location; the absolute rate of change in $hB_{\rm max}^2$ increases with the mass influx rate.  

Now, B11 assumes that $\mdot \propto hB_{\rm max}^2$; consequently, an argument similar to that above leads him to conclude that the mass pile-up at any such location is a runaway process: a net influx of material reduces $hB_{\rm max}^2$, which in turn yields (with his formula) a smaller $\mdot$ out of this location and hence even higher net influx, and so on.  The general formula we use for the accretion rate, however, paints a more complex picture.  Since $hB_{\rm max}^2$ (and hence also $r^2hB_{\rm max}^2$, for a given $r$) decreases faster for higher influx rates, we have in general: $\partial(r^2hB_{\rm max}^2)/\partial{t} = - f(\partial{\mdot}/\partial{r})$, where $f$ is some positive function that captures the details of this evolution.  With $\mdot \propto \partial(r^2 h B_{\rm max}^2)/\partial{r}$, we finally get: $\partial{\mdot}/\partial{t} \propto - \partial[f(\partial{\mdot}/\partial{r})]/\partial{r}$.  If we {\it ignore} the outer differential, as B11 essentially does, then indeed $\mdot$ diminishes with time ($\partial{\mdot}/\partial{t} < 0$) for a net mass influx ($\partial{\mdot}/\partial{r} > 0$), leading to a runaway pile-up.  The general equation, though, shows that whether $\mdot$ is enhanced or reduced with time depends on the {\it radial variation} of the mass influx rate $\partial{\mdot}/\partial{r}$; the process only runs away if the latter increases with $r$, which is not necessarily true even if $\partial{\mdot}/\partial{r} > 0$.  Analogous arguments can be made for the slightly different phenomenologies that apply when the active layer extends to the midplane, and/or there is a net mass outflux instead of influx.  In summary, the final outcome may be runaway mass pile-up in some regions, runaway evacuation in others, and oscillations about (or towards) a steady-state in yet others, depending on the precise variations (both first and second-order) of the accretion rate with radius (and the specifics of the conditions in the active layer, which determine the function $f$).  Further investigation of this requires detailed time-dependent simulations, beyond the scope of this paper.           

\noindent {\it (iv)} Whatever the final, or time-dependent, structure of the disk might be, the above discussion shows that it is almost certain to comprise regions of both over- and underdensity; achieving steady-state with a surface density that decreases monotonically with radius appears highly unlikely.  On the one hand, this implies that the formation of such disk structures, currently attributed to either sculpting by planets or photoevaporation, may be a natural outcome of the MRI-driven accretion itself.  On the other hand, this may enhance the former two processes: photoevaporation will be more efficient in rarefied (more optically thin) regions, while grain growth can be accelerated in overdense regions (due to both the increased density, as well as heightened trapping of grains due to the pressure gradient reversal caused by the local density enhancement, similar to the situation envisaged by Ida \& Lin (2008) near the ice-line).  Finally, both disk gaps and pressure gradient reversals due to overdensities can halt planetary migration \citep{2012MNRAS.422L..82A, 2008ApJ...685..584I}, possibly making Type I migration timescales long enough to be consistent with observations.      

\noindent {\it (v)} In general, our inferred accretion rates increase outwards with radius, before falling off again in the outer disk, broadly consistent with the results of B11 (except that our $\mdot$ often reverses direction in the outer disk as well).  The mean {\it empirical} rate of infall onto young stars -- $\sim 10^{-8}$\,$\msun$\,yr$^{-1}$ for solar type stars and $10^{-10}$\,$\msun$\,yr$^{-1}$ for very low mass ones \citep[e.g.,][]{2005ApJ...626..498M, 2005ApJ...625..906M} -- is achieved in the innermost parts of our disks for dust-to-gas mass ratios of $\leq 10^{-5}$ for 0.7\,$\msun$ and $\leq 10^{-3}$ for 0.1\,$\msun$ respectively.  Thus a factor of $\sim$10--1000 depletion in the mass sequestered in well-mixed small grains, relative to the standard ISM ratio of $10^{-2}$, appears necessary to reproduce the observed stellar $\mdot$.  Given that grain growth up to millimeter sizes seems common in these stars even by $\sim$1\,Myr \citep{2005ApJ...631.1134A, 2007ApJ...671.1800A, 2010AA...521A..66R, 2010AA...512A..15R, mohantysub}, this appears plausible: since the mass of a grain rises as the cube of its radius, very few large grains are necessary to significantly depress the mass locked up in small ones (e.g., for a 1000-fold depletion in the total 0.1$\mu$m grain mass, only one in a billion grains in the depleted disk needs to be millimeter-sized).  These levels of depletion in small grains are also consistent with mid-infrared observations of T Tauri disks \citep[e.g.,][]{2006ApJS..165..568F}. 

In this context, we point out that the $\mdot$ we derive using B11's formula are 10--20 times larger than he finds, for the case of a solar-type star (0.7\,$\msun$) with 0.1\,$\mu$m grains and a dust-to-gas ratio of $10^{-2}$; for the grain-free case, our values with his formula are up to a 100 times larger than his.  In both cases, this is because our $B_{\rm max}$ are larger, because of higher electron abundances.  In the first instance, this is due to our more detailed treatment of the X-ray flux and Compton scattering, as well as some differences in the chemistry and surface density (see discussion in \S6.2.1), leading to 3--5$\times$ larger $x_e$ (and thus $B_{\rm max}$).  In the grain-free case, as discussed at length in \S6.1, our $x_e$ are considerably higher, due to the difference in our chemical networks.  

\noindent {\it (vi)}  Finally, in spite of our somewhat larger $B_{\rm max}$ in the presence of grains, compared to the results of B11, our field strengths at 50--100\,AU are still $<$10\,mG, for 0.7\,$\msun$ with dust-to-gas ratios $10^{-2}$--$10^{-5}$ (and lower yet for 0.1\,$\msun$).  We therefore agree with B11 that the current {\it non}-detection of polarized emission from grains aligned with the field is consistent with {\it active} MRI.  Specifically, SMA observations by \citet{2009ApJ...704.1204H}, with angular resolution corresponding to 50--100\,AU, have failed to detect such emission at mm-wavelengths from two solar-type classical T Tauri stars.  Their models suggest that most of  this emission is expected from 10--100\,$\mu$m grains, for which the minimum field strength for grain alignment is 10--100\,mG.  Our results, however, show that fields this strong would disrupt the MRI in the outer disk in the presence of even a tiny fraction of very small grains; the non-detection, implying weaker fields, is thus compatible with active MRI.    

\section{Conclusions and Future Work}
We have presented numerical simulations of X-ray driven MRI in disks
around both solar-type and very low mass stars, including the the
effects of ambipolar diffusion, in the presence of 0.1\,$\mu$m grains.
In his seminal study, B11 undertook an analogous investigation of
solar-type stars.  In addition to extending his work to the very low
mass stellar domain, we have incorporated several improvements to his
treatment: a more sophisticated analysis of the X-ray ionization; a
generalized formulation of the accretion rate; an investigation of
various degrees of grain depletion between the limiting cases of a
standard ISM dust abundance and no dust at all; and an examination of
the active layer properties over the entire disk from 0.1 to 100\,AU.
Our choice of the field strength at every radius is governed by the
assumption that the MRI generates a magnetic field of the strength
that maximizes the height-integrated accretion stress, which is very close to the maximum possible field strength that still permits active MRI (the same choice
as B11 makes, but for a different physical reason).  Our main
conclusions can be summarized thus:

\noindent $\bullet$  Ambipolar diffusion is the primary factor determining the thickness and location of the MRI-active layer around both solar-type and very low mass stars, and quenches the MRI over most of the disk: the active region constitutes only 5--10\% of the total disk mass in all cases.  Previous suggestions that reduced surface density in low-mass disks (around less massive stars) may cause the entire disk to become active \citep[e.g.,][]{2006ApJ...648..484H} are invalid, precisely because ambipolar diffusion becomes stronger at lower densities.  

\noindent $\bullet$ Additionally, small well-mixed grains strongly suppress the MRI.  Achieving the empirical accretion rates in solar-type and very low mass stars requires that the total mass in such grains be depleted (via grain growth and/or settling) by a factor of 10--1000, relative to the standard dust-to-gas mass ratio of $10^{-2}$.  This degree of depletion is not implausible, given the rapid grain growth observed in classical T Tauri stars, and is consistent with the range of depletion suggested by mid-infrared observations of these disks \citep{2006ApJS..165..568F}.

\noindent $\bullet$ The accretion rates are radially variable, and may even change sign (direction) at some locations.  This can produce both runaway pile-up or evacuation of material at some radii, and oscillations about a steady-state in others.  Radially, the disk is thus expected to comprise regions of both over- and underdensity. These can accelerate photoevaporation as well as grain coagulation and planetesimal formation, and reduce the efficiency of Type I migration.

\noindent $\bullet$ The current non-detection of polarized emission from field-aligned grains in the outer disk is consistent with active MRI in these regions.  

Various questions still remain open, however.  Some of the most critical, offering profitable avenues for future investigation, are as follows.  

\noindent $\bullet$ {\it Chemistry}: As discussed above, we find (in agreement with B11 and PC11) that a very substantial depletion of small grains is required for X-ray driven MRI to reproduce the observed accretion rates in classical T Tauri stars. However, it is precisely at such low grain abundances that the derived MRI accretion rate becomes sensitive to the particular chemical network adopted (as indicated by the significant disparity between our $\mdot$ and that of B11, in the dust-free case).  Such severe small-grain depletion may also favour the dominance of atomic hydrogen over molecular, contrary to our and B11's assumption (as discussed in \S4.2).  Thus, a more detailed investigation of chemical networks and hydrogen recombination channels at very low grain abundances is essential to confirm (or refute) the importance of X-ray driven MRI for disk accretion.

\noindent $\bullet$ {\it PAHs}: In an elegant semi-analytic study, PC11 showed that PAHs -- macromolecular sized dust particles -- may pose an even graver threat to X-ray-driven MRI than the small grains considered here.  This is because the efficiency with which dust grains soak up electrons (and hence also the rate of ion recombinations on grain surfaces) increases with the total grain surface area, and for a fixed dust mass, the total area rises in inverse proportion to the grain radius.  PC11 consider PAHs of radius 6\,\r{A}, which would enhance the area available for adsorption and recombination by a factor of 200 compared to our 0.1\,$\mu$m grains, and thus, prima facie, enormously suppress the MRI.  However, PC11 neglect the competing fact that grains are also better tied to field lines as they become smaller: not because of a reduced grain mass (which is irrelevant as long as it far exceeds that of a neutral particle), but because of a reduced collision frequency with neutrals.  Specifically, from equation (19), the ratio of the Hall parameters for PAHs and ions (a measure of their relative attachments to the field) is $\beta_{\rm PAH}/\beta_i \approx (|Z_{\rm PAH}|<\sigma v>_i)/(|Z_i|<\sigma v>_{\rm PAH})$.  At a fiducial disk temperature of $\sim$300\,K, corresponding to around 1\,AU for a solar-type star, equation (26) implies that the collisional rate coefficient for 6\,\r{A} PAH grains becomes $<\sigma v>_{\rm PAH} \,\sim 10^{-9}$\,cm$^3$\,s$^{-1}$, comparable to that for ions (equation (25), for both molecular and metallic ions).  Thus, even if most small PAHs carry a single charge (as PC11 find), they are still just as bound to the field as the ions (which are predominantly singly-charged as well), at this temperature.  In cooler parts of the disk, small PAHs are even better tied to the field than ions are (since the rate coefficient for grains goes as $\sqrt{T}$ while that for ions is constant); the same is true if the PAHs are multiply-charged.  In this situation, PAHs complement ions in tying neutrals to the field through collisions, thereby reducing ambipolar diffusion.  If this effect overcomes the increased recombination these particles facilitate, then PAHs can {\it promote} the MRI instead of hindering it.  This very promising volte-face by PAHs richly deserves further study.

\noindent $\bullet$ {\it FUV-driven MRI}:  Following the work of PC11, \citet{2011ApJ...735....8P} showed that the MRI can thrive even in the presence of PAHs if the ionization is driven by stellar FUV photons instead of X-rays: while the FUV radiation does not penetrate as far into the disk, it ionizes carbon and sulphur atoms to yield electron and ion fractional abundances orders of magnitude higher than achieved with X-rays, completely dwarfing PAH effects.  At radii $\gtrsim$\,10\,AU, the maximum inferred $\mdot$ in this case is comparable to the mean observed rate of $10^{-8}$\,$\msun$\,yr$^{-1}$ for solar-type stars, though ambipolar diffusion again causes the rate to fall well below the observed mean in the inner disk regions.  Thus FUV ionization may explain the empirical accretion rates of classical T Tauri stars, at least beyond 10\,AU, without having to invoke significant depletion of small grains.  However, it is unclear how much of the FUV radiation actually reaches the disk surface: in accreting stars, these photons are mainly generated in the accretion shocks on the stellar surface, and should be strongly attenuated on their way to the disk by the surrounding infalling accretion columns.  An analysis accounting for this effect is required to better evaluate the viability of FUV-driven MRI.  

\noindent $\bullet$ {\it $\mdot$ variability and time-dependence}: We find the accretion rates to be radially variable; this is a general feature of layered accretion (e.g., \citet{1996ApJ...457..355G}).  In addition, we see that this not only leads to a buildup of mass in the inner disk \citep[e.g.,][B11]{1996ApJ...457..355G}, but can cause either a pileup or evacuation of material at several locations, depending on how $\mdot$ changes with radius.  While the importance of mass pileup for non-steady accretion and disk instabilities is recognized, the effects of accretion-related radial oscillations in density on planet formation and migration have been little studied (but see \citet{2008ApJ...685..584I} for a discussion of viscosity-related overdensity near the ice-line, and \citet{2012MNRAS.422L..82A} for the effect of photoevaporation-driven gaps on migration).  Doing so first requires describing these density variations in more detail, through time-dependent MRI accretion simulations in the future.  Such dynamic simulations are also required to better treat the time-dependent effects of grain growth and settling \citep[e.g.,][]{2010ApJ...708..188T}.    

\noindent $\bullet$ {\it non-ideal MHD}: Finally our investigation, like that of B11, is based on the latest numerical simulations of non-ideal MHD, which still have a number of important uncertainties.  Two in particular are worth mentioning.  The first is that our ambipolar diffusion criterion for the MRI, equation (4), originates in the unstratified calculations by \citet{2011ApJ...736..144B}; whether this continues to hold in the stratified domain remains to be seen.  Second, we have ignored the effects of Hall diffusion on the MRI, guided by the non-linear simulations of \citet{2002ApJ...577..534S}.  As \citet{2011arXiv1103.3562W} point out, however, these calculations do not probe the ``deep'' Hall regime, where $|\eta_H| > \eta_P > v_A^2/\Omega$.  Wardle \& Salmeron's {\it linear} analysis, with vertical magnetic fields, suggests that Hall diffusion in this regime can change the column density of the active layer by an order of magnitude or more (either increasing or decreasing the column depending on whether the field is parallel or antiparallel to the rotation axis).  Non-linear simulations are now essential to unravel the true role of Hall diffusivity here.  The bottom line is that analyses such as ours depend perforce on the results of non-ideal MHD simulations, and advances in the latter are vital to further improve our understanding of the viability and importance of the MRI for protoplanetary disk accretion.  

\bigskip
\bigskip

{\it Acknowledgements:} We would like to thank Xue-Ning Bai, Daniel
Perez-Becker, Eugene Chiang and Raquel Salmeron for invigorating
discussions and very useful insights.  We would also like to thank the anonymous referee for a very thoughtful report, which helped improve the paper considerably.  S.M.\,is very grateful to the
{\it International Summer Institute for Modeling in Astrophysics}
(ISIMA) for affording him the time, research environment and
interactions necessary to take this work forward, and acknowledges the
funding support of the STFC grant ST/H00307X/1.  N.J.T.\ was
a guest at MPIA Heidelberg when this work began, and especially thanks
Thomas Henning for pointing out the importance of the topic.  N.J.T.\
was supported by the Alexander von Humboldt Foundation under a
Fellowship for Experienced Researchers and by the the NASA Origins of
Solar Systems program.  He is employed at the Jet Propulsion
Laboratory, California Institute of Technology, under a contract with
NASA.

\bibliographystyle{apj}
\bibliography{paper}
\clearpage

\begin{figure}
\includegraphics[scale=0.9]{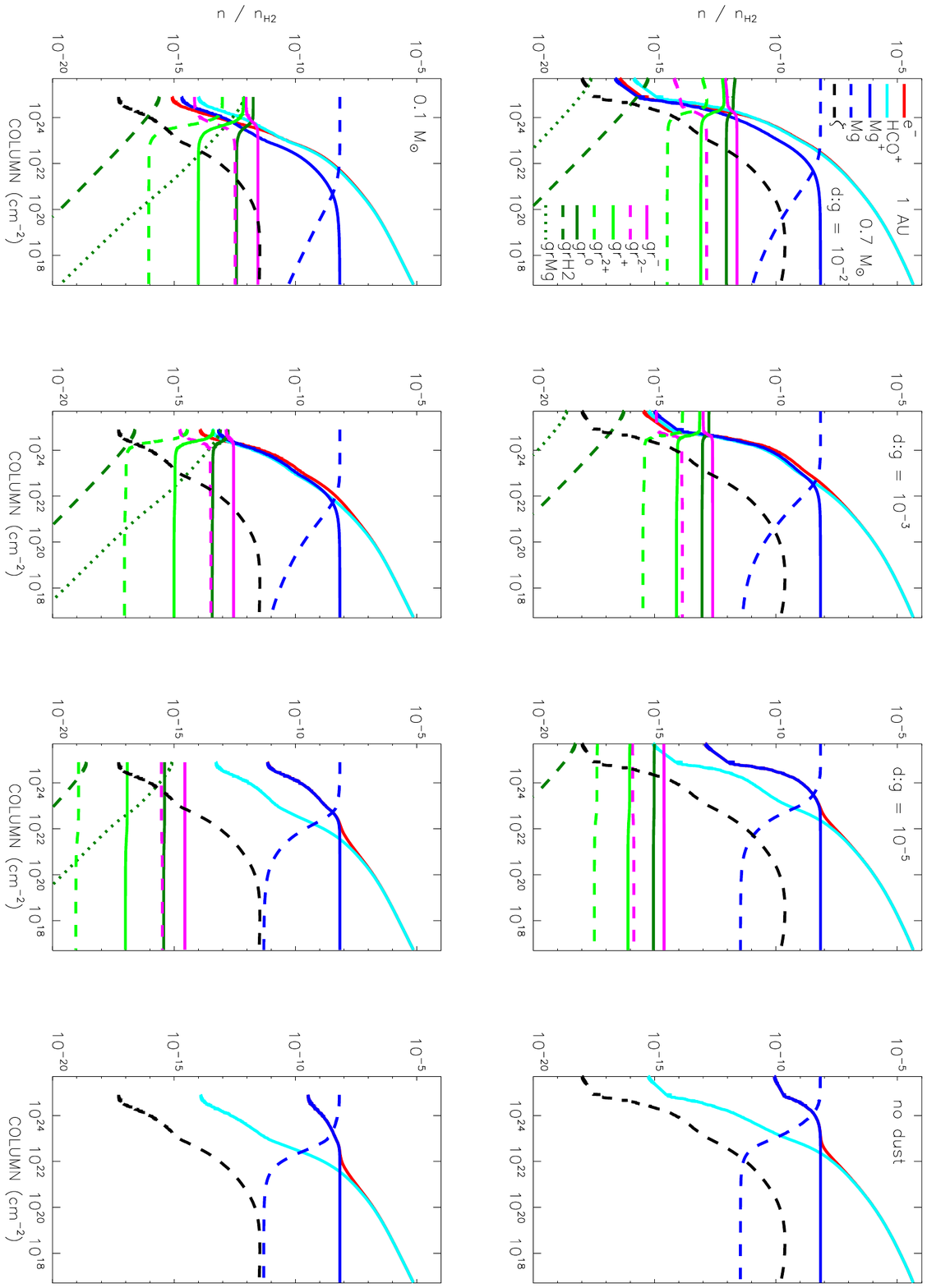}
\end{figure}
\begin{figure}
\caption{Ionization rate ({\it black dashed line}) and fractional abundances of the various species in our chemical network ({\it colored lines}, specified in key in top left panel), as a function of column density at 1\,AU, for various dust-to-gas ratios and two stellar masses.  {\it Top}: 0.7\,$\msun$. {\it Bottom}: 0.1\,$\msun$.  {\it Left to Right}: dust-to-gas ratio = $10^{-2}$, $10^{-3}$, $10^{-5}$, 0.  See \S6.1.}
\end{figure}

\begin{figure}
\includegraphics[scale=0.9]{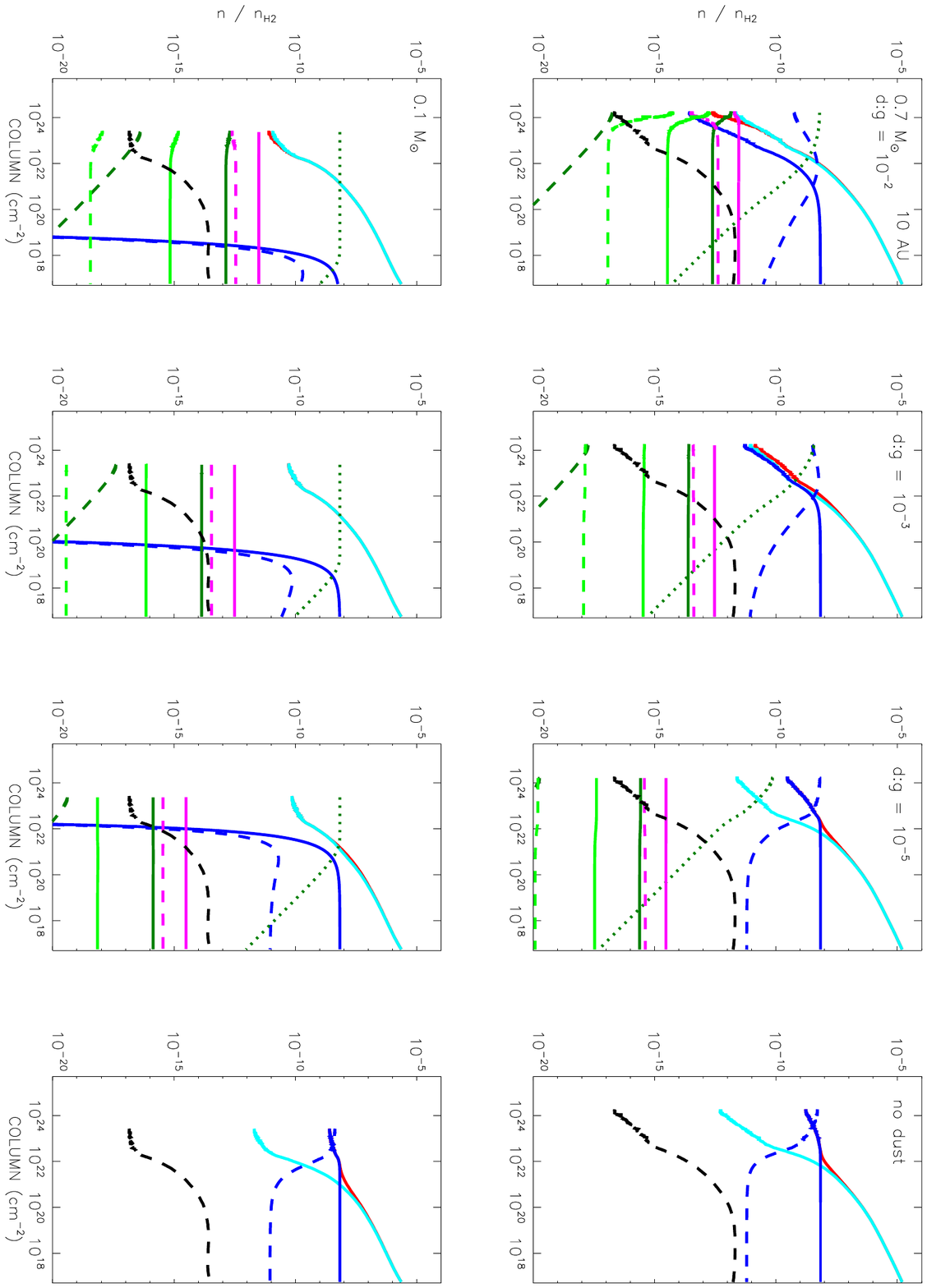}
\end{figure}
\begin{figure}
\caption{Same as Fig.\,1, but at 10\,AU.}
\end{figure}

\begin{figure}
\includegraphics[scale=0.5, angle=90]{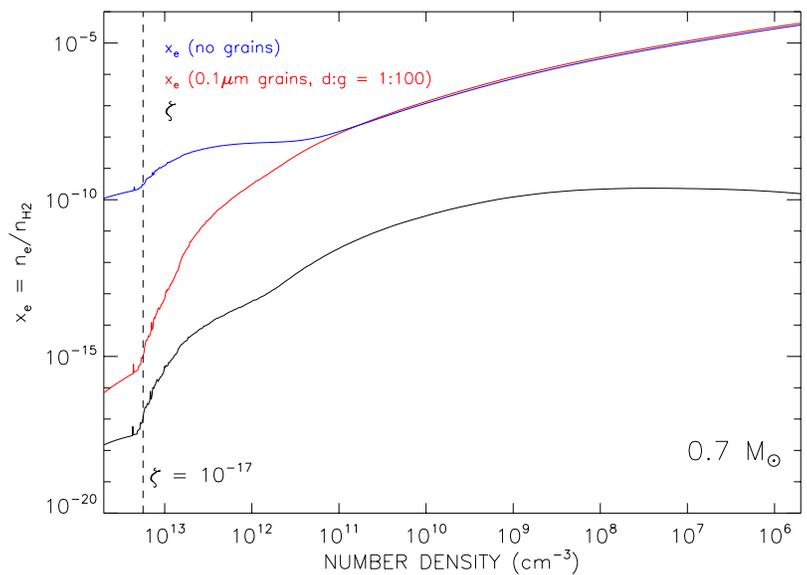}
\includegraphics[scale=0.5, angle=90]{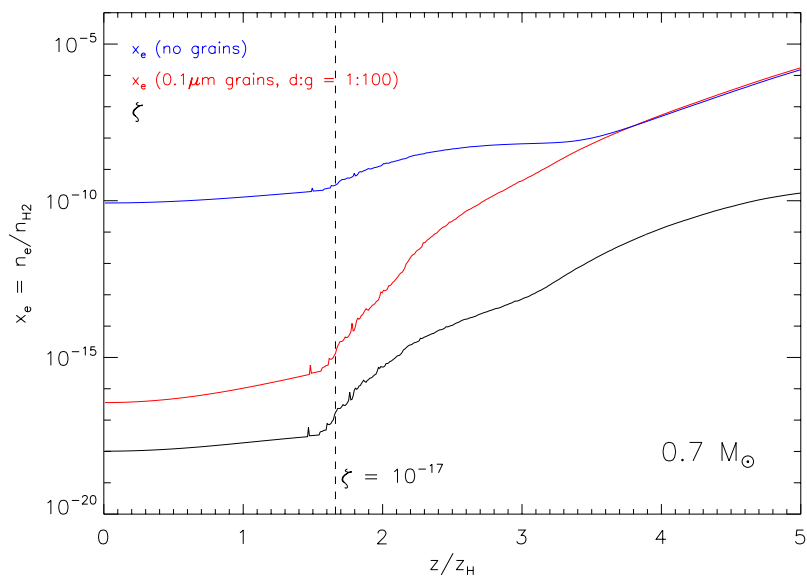}
\caption{Ionization rate $\zeta$ ({\it black line}) and fractional electron abundance $x_e$ ({\it red}: with dust-to-gas ratio = $10^{-2}$, {\it blue}: no-dust), at 1\,AU for 0.7\,$\msun$, as a function of number density ({\it top panel}) and height ({\it bottom panel}).  The {\it dashed vertical line} shows $\zeta = 10^{-17}$.  See \S6.1.  }
\end{figure}

\begin{figure}
\includegraphics[scale=0.55, angle=90]{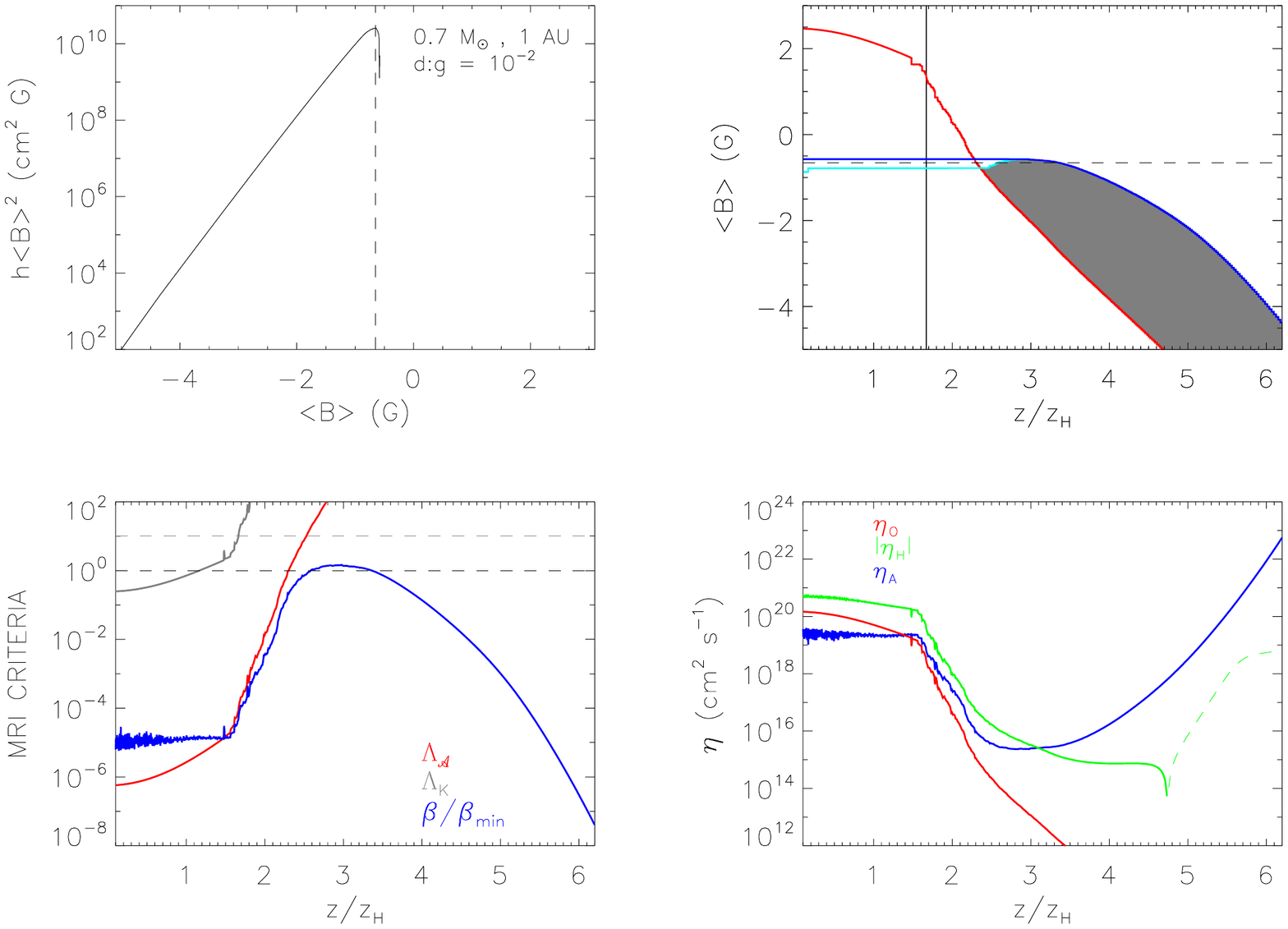}
\includegraphics[scale=0.55, angle=90]{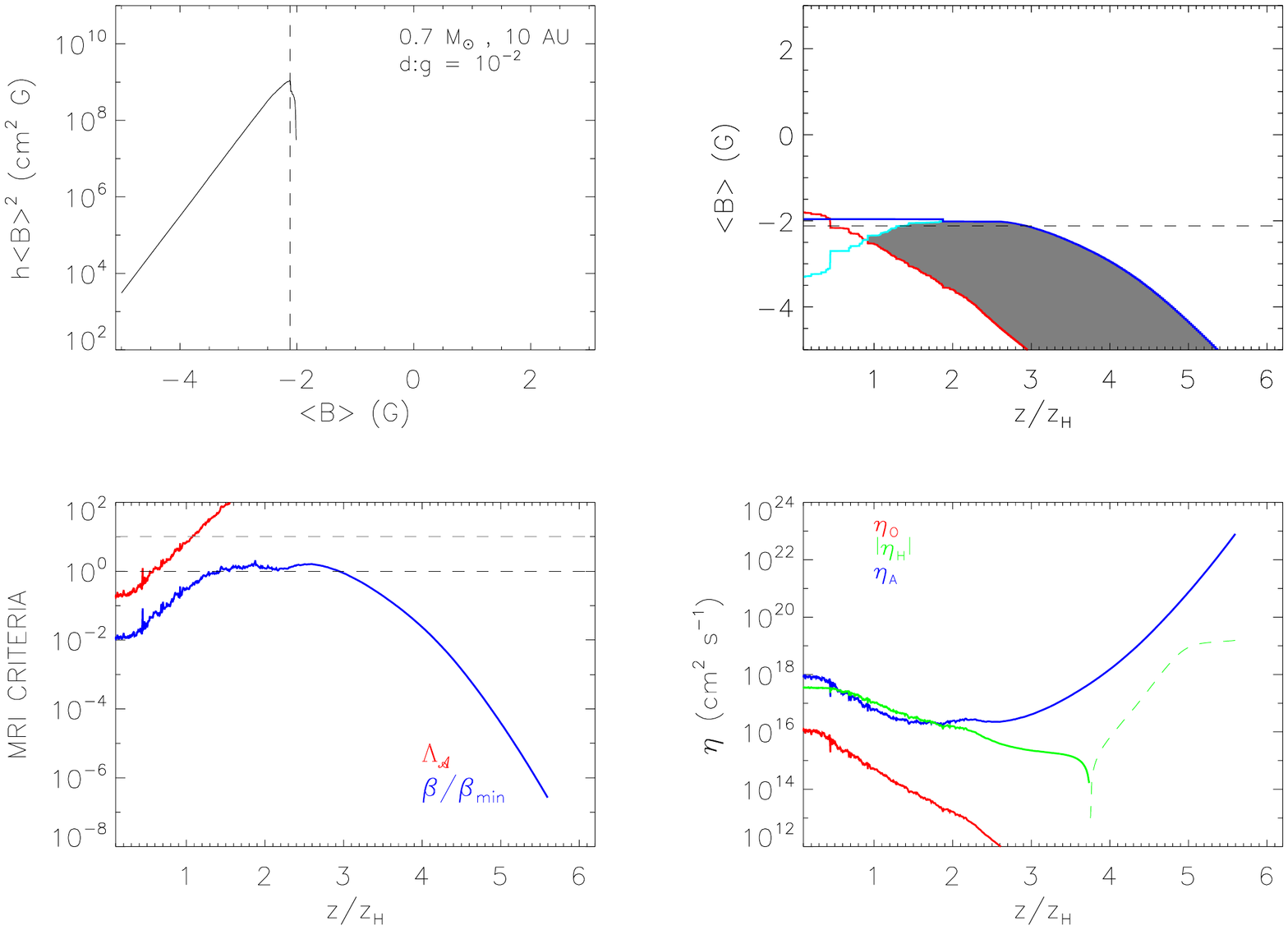}
\end{figure}
\begin{figure}
\caption{Various field-dependent quantities, for 0.7\,$\msun$.  {\it Top plot} (a): At 1\,AU.  {\it Bottom plot} (b): At 10\,AU.  Within each plot, we show:- {\it Top left}: $hB^2$ as a function of field strength $B$.  The {\it dashed vertical line} marks the field strength at which $hB^2$ is maximum; note that this is very close to the maximum $B$ allowed (which is the end of the curve).  We set $B_{\rm max}$ equal to the $B$ at which $hB^2$ is maximized.  {\it Top right}: MRI-active layer ({\it grey region}) as a function of field strength and vertical height.  $\Lambda_{\mathcal A} > 1$ everywhere to the right of the {\it red curve}; $\Lambda_{K} > 10$ everywhere to the right of the {\it grey vertical line}; and $\beta/\beta_{\rm min} > 1$ everywhere to the left of the {\it blue curve} and right of the {\it aqua curve}.  The {\it horizontal dashed line} marks $B = B_{\rm max}$; note that at this field strength, both the bottom and top of the active region are set by the ambipolar condition (blue and aqua curves).  {\it Bottom left}: $\Lambda_{\mathcal A}$ ({\it red}), $\Lambda_K$ ({\it grey}) and $\beta/\beta_{\rm min}$ ({\it blue}) as a function of height, for $B = B_{\rm max}$. $\Lambda_{\mathcal A} > 1$ and $\beta/\beta_{\rm min} > 1$ above the {\it black dashed line}, and $\Lambda_{\mathcal K} > 10$ above the {\it grey dashed line}.  {\it Bottom right}:  Various resistivities as a function of height, for $B = B_{\rm max}$.  $\eta_O$ in {\it red}, $|\eta_H|$ in {\it green} ({\it solid} and {\it dashed}: positive and negative $\eta_H$ respectively), $\eta_A$ in {\it blue}.  See \S6.2.1.               }
\end{figure}

\begin{figure}
\includegraphics[scale=0.7]{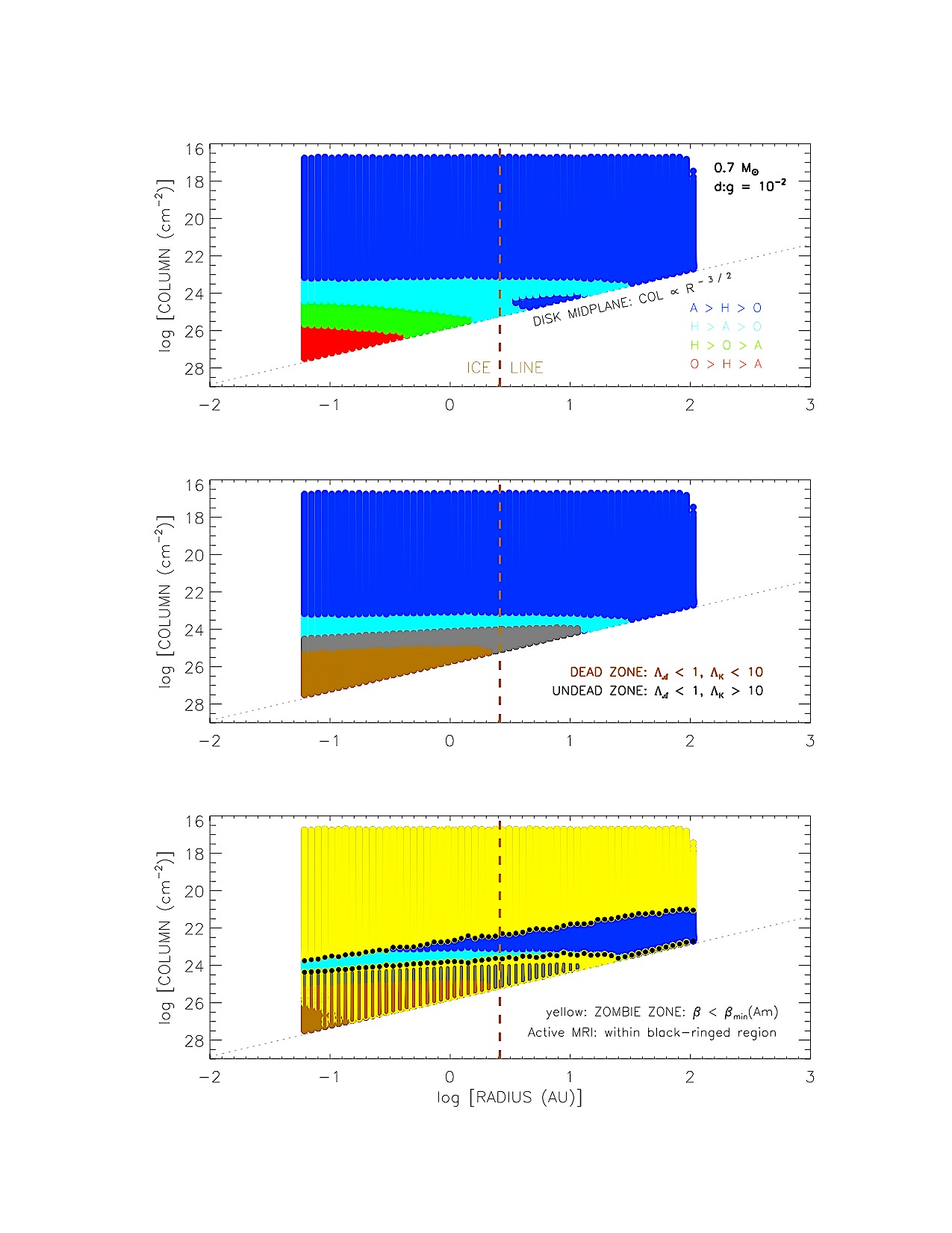}
\end{figure}
\begin{figure}
\caption{Various resistivity and activity zones over the entire disk, for 0.7\,$\msun$ and dust-to-gas ratio = $10^{-2}$.  {\it Top panel}: Dominant resistivities (colors specified in plot key).  {\it Middle panel}: Same, with Dead Zone ({\it brown}) and Undead Zone ({\it grey}) overplotted.  {\it Bottom panel}: Same, with Zombie Zone ({\it yellow}) overplotted as well.  Where the Zombie Zone overlaps with the dead and Undead Zones, the latter are shown as a picket-fence.  The MRI-active zone lies within the {\it black-ringed} region.  Note that the bottom of the active zone is always set by the Zombie Zone (ambipolar condition), in this and all following plots (not obvious at the smallest radii at this plotting resolution, but nevertheless true).  See \S6.2.2. }
\end{figure}

\begin{figure}
\includegraphics[scale=0.7]{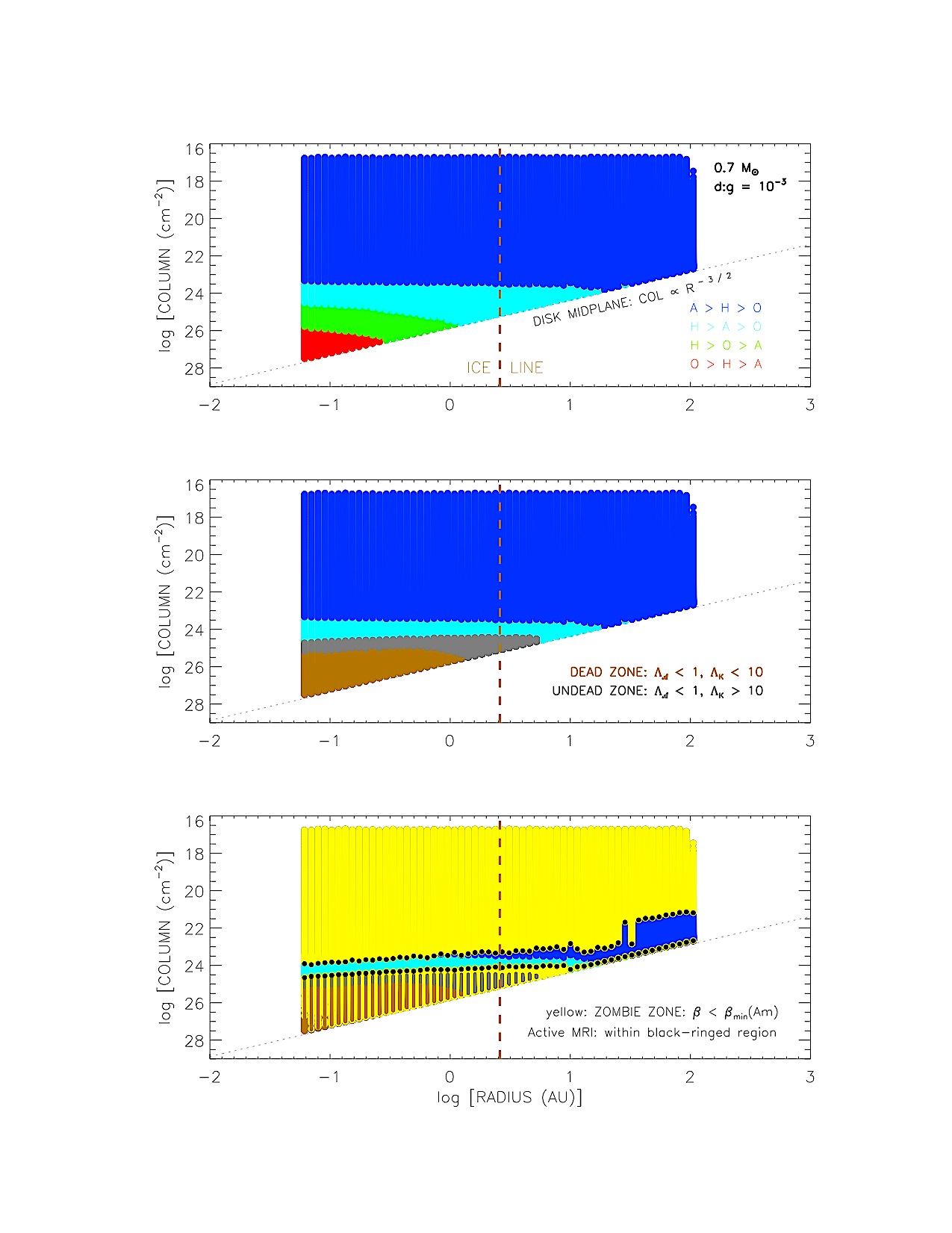}
\caption{Same as Fig.\,5, but for 0.7\,$\msun$ and dust-to-gas ratio = $10^{-3}$.}
\end{figure}

\begin{figure}
\includegraphics[scale=0.7]{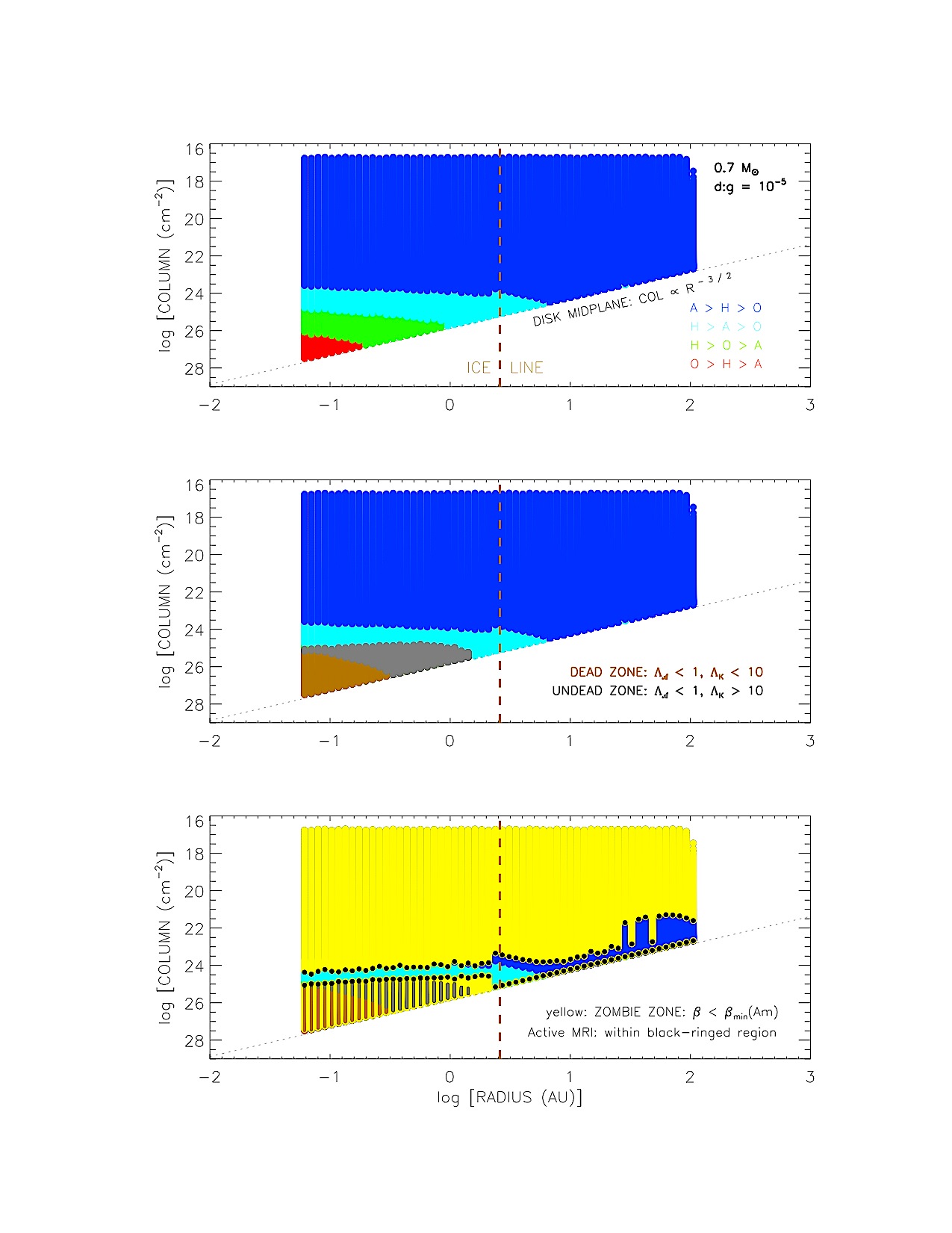}
\caption{Same as Fig.\,5, but for 0.7\,$\msun$ and dust-to-gas ratio = $10^{-5}$.}
\end{figure}

\begin{figure}
\includegraphics[scale=0.7]{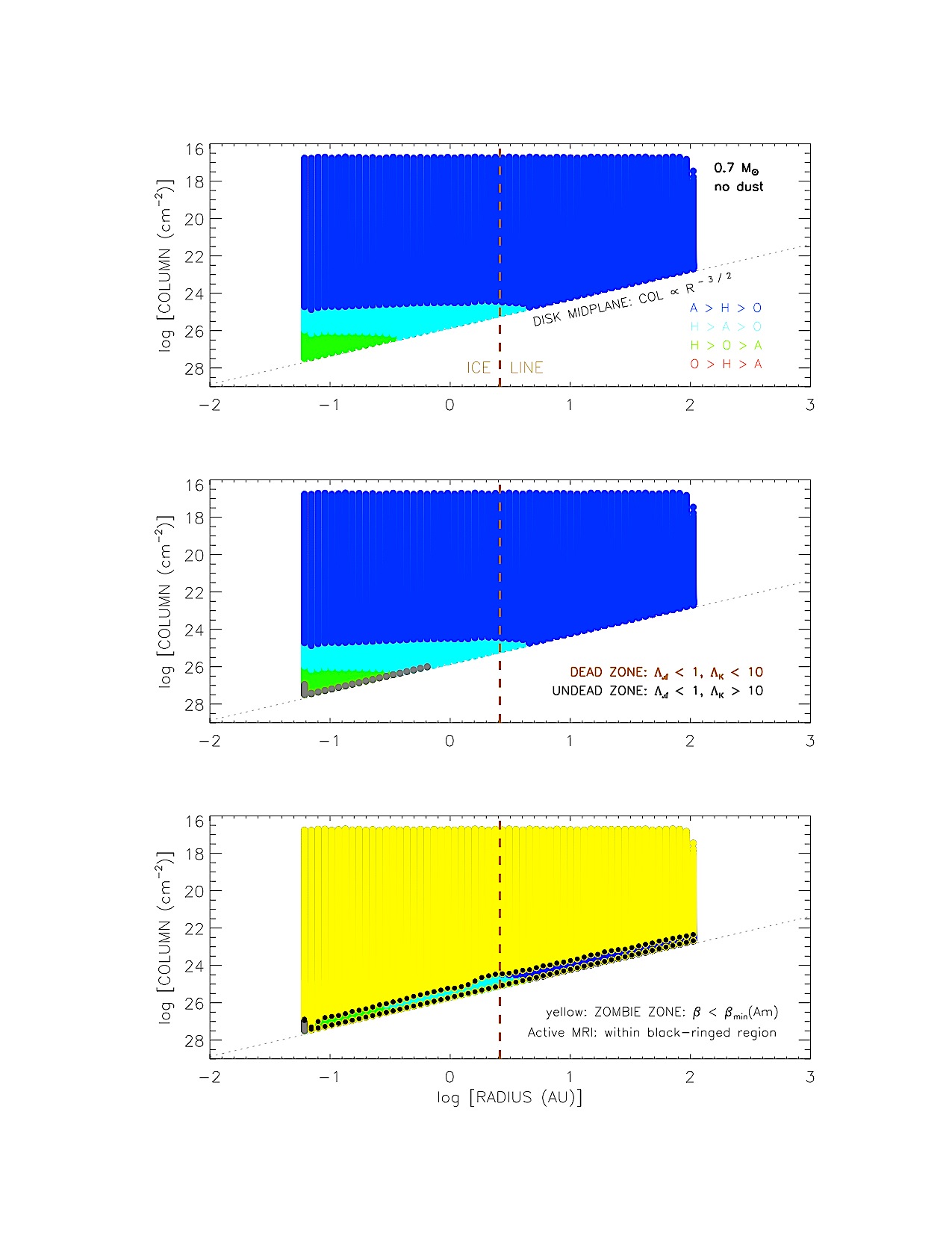}
\caption{Same as Fig.\,5, but for 0.7\,$\msun$ and dust-to-gas ratio = 0 (no dust).}
\end{figure}

\begin{figure}
\includegraphics[scale=0.7]{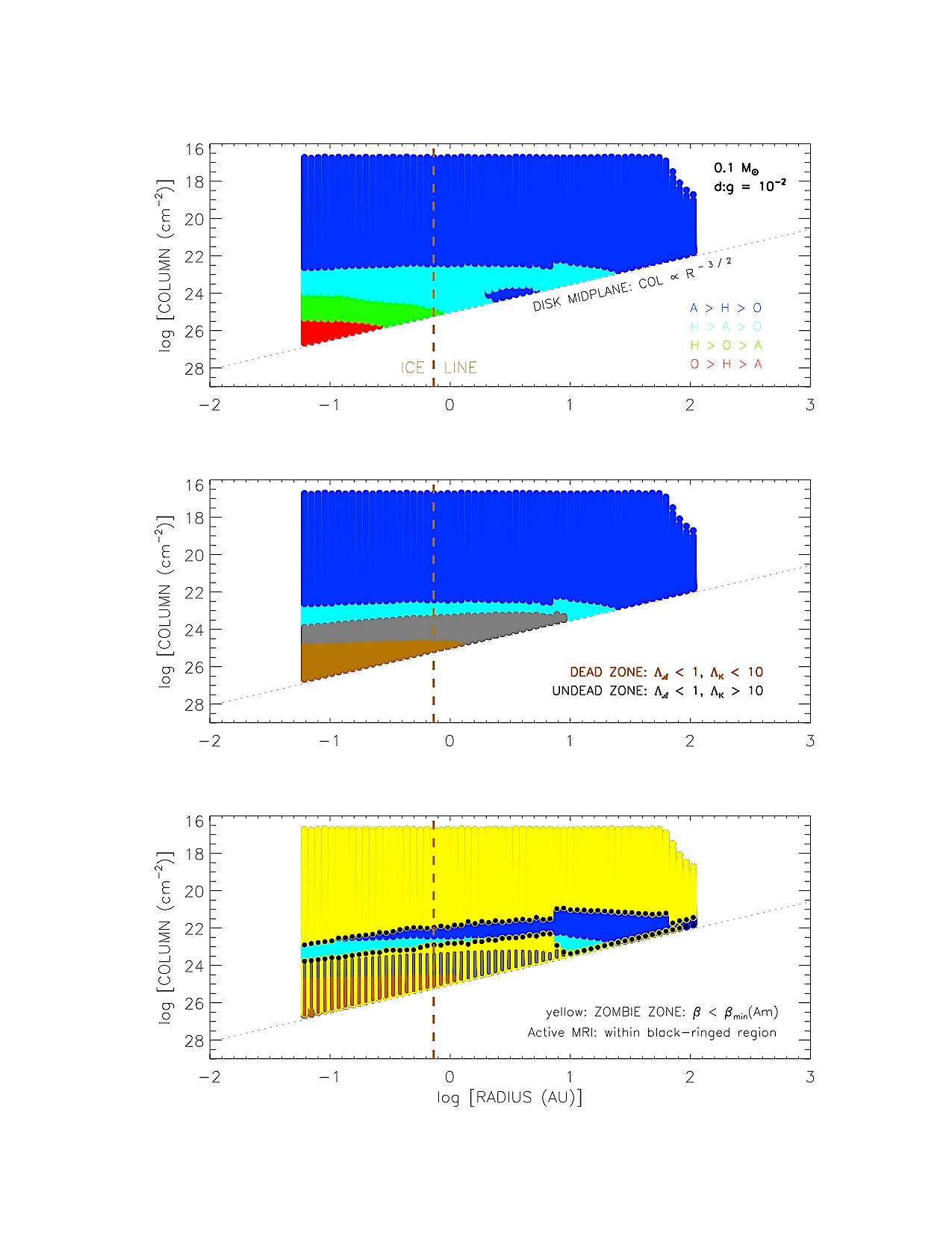}
\caption{Same as Fig.\,5, but for 0.1\,$\msun$ and dust-to-gas ratio = $10^{-2}$. }
\end{figure}

\begin{figure}
\includegraphics[scale=0.7]{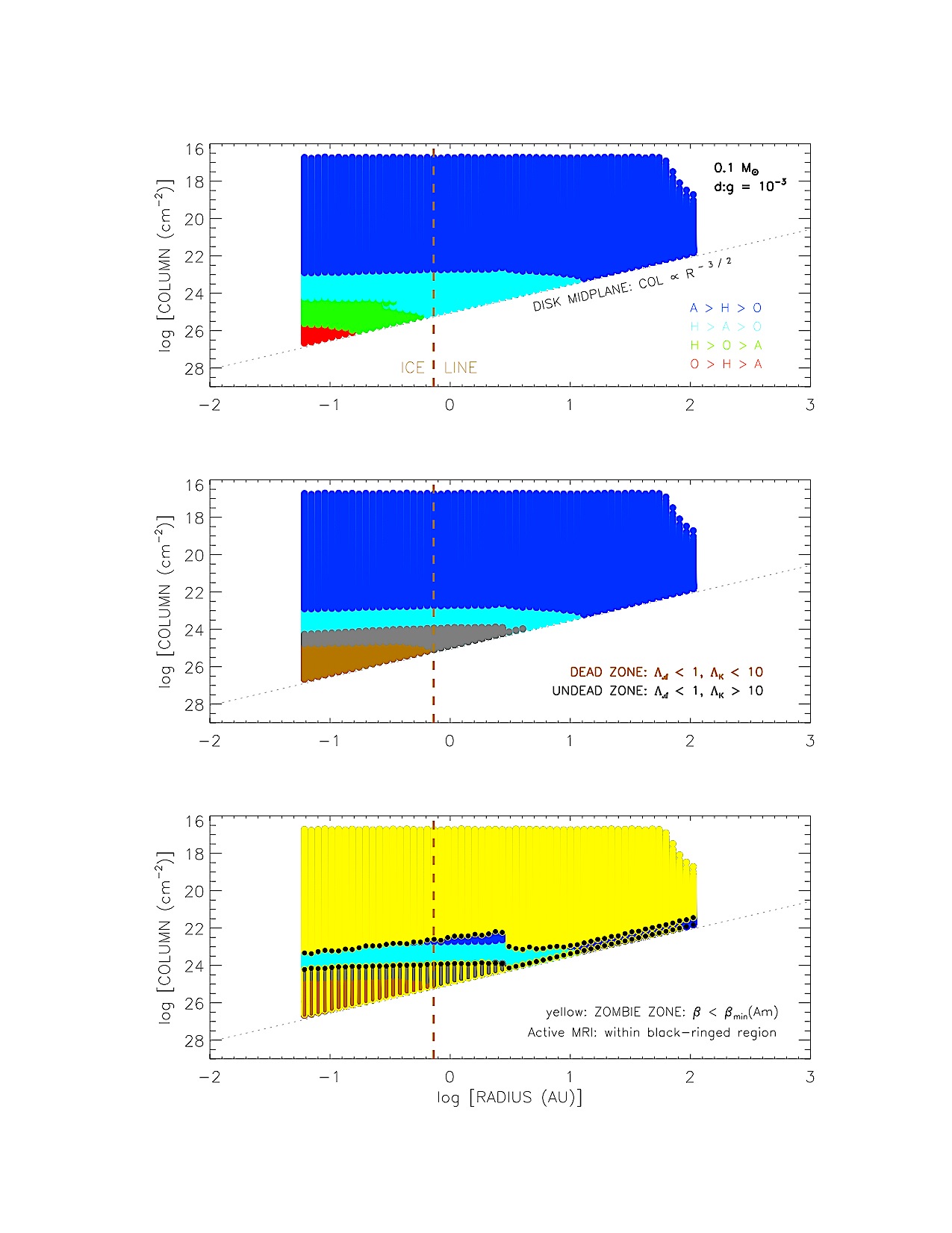}
\caption{Same as Fig.\,5, but for 0.1\,$\msun$ and dust-to-gas ratio = $10^{-3}$.}
\end{figure}

\begin{figure}
\includegraphics[scale=0.7]{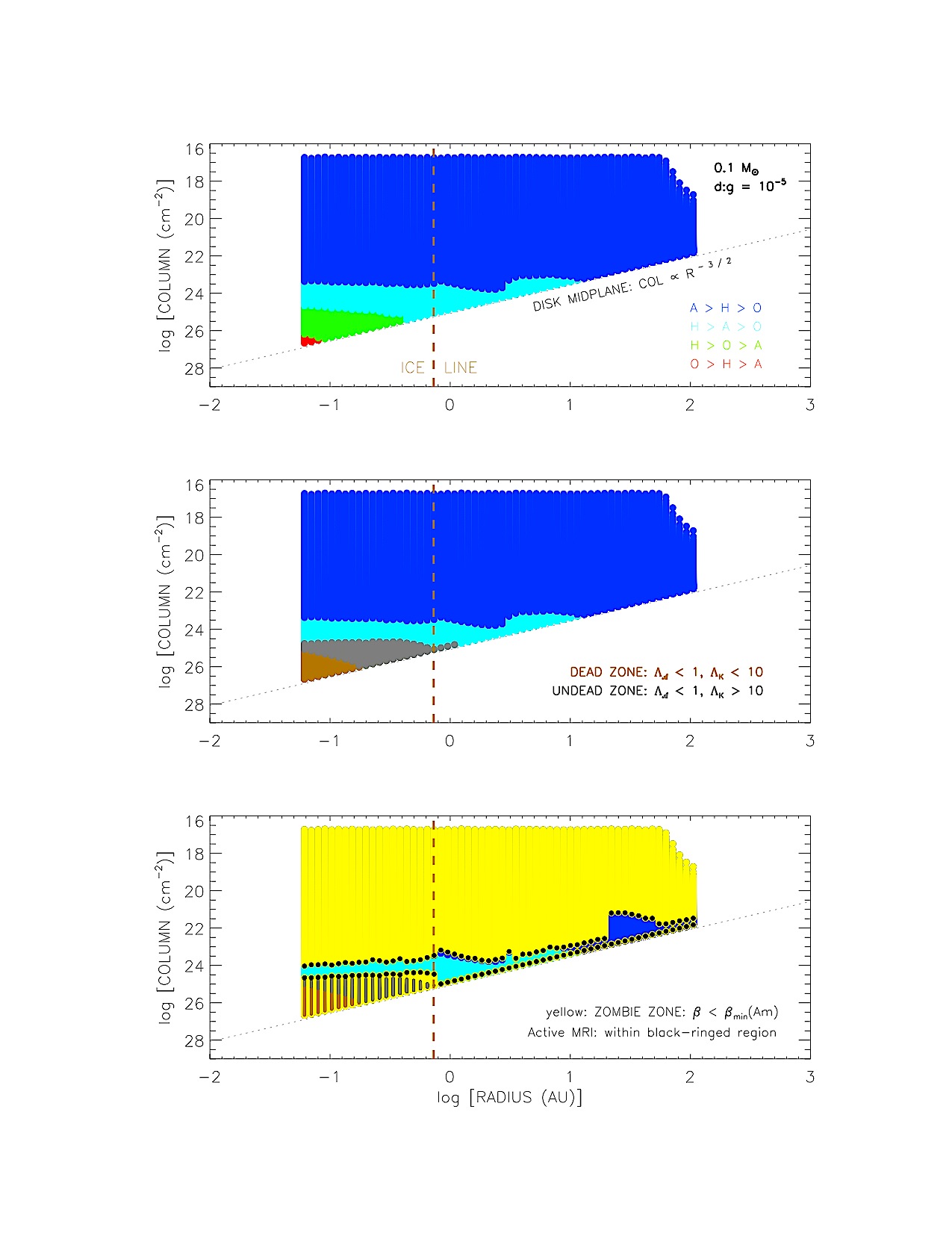}
\caption{Same as Fig.\,5, but for 0.1\,$\msun$ and dust-to-gas ratio = $10^{-5}$.}
\end{figure}

\begin{figure}
\includegraphics[scale=0.7]{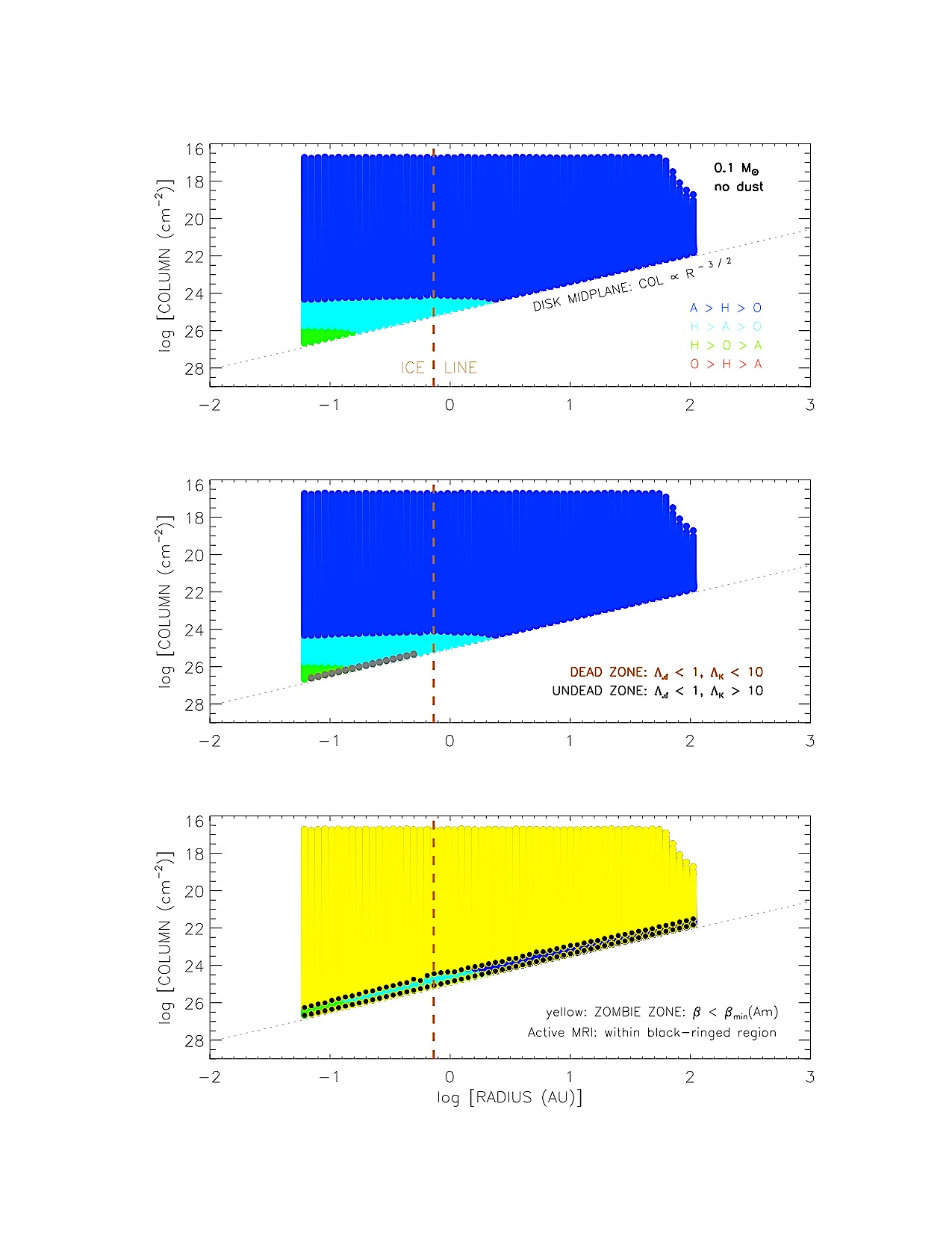}
\caption{Same as Fig.\,5, but for 0.1\,$\msun$ and dust-to-gas ratio = 0 (no dust).}
\end{figure}

\begin{figure}
\includegraphics[scale=0.7, angle=90]{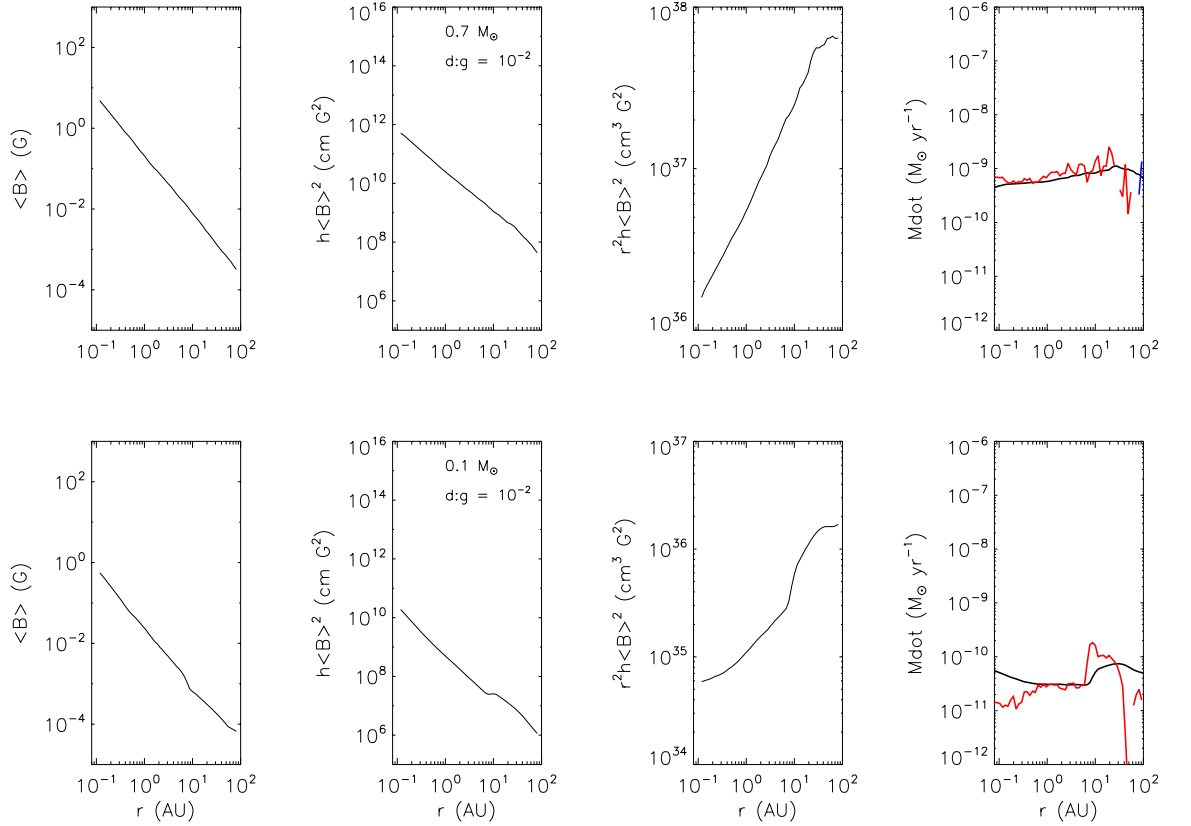}
\caption{Various quantities relevant to the accretion rate, plotted as a function of radius, for dust-to-gas ratio = $10^{-2}$.  {\it Top panels}: 0.7\,$\msun$.  {\it Bottom panels}: 0.1\,$\msun$.  {\it Left to right}: $B_{\rm max}$; $hB_{\rm max}^2$, which sets the accretion rate (modulo $\Omega$) using the formula of Bai (2011); $r^2 h B_{\rm max}^2$, the {\it slope} of which sets the accretion rate (modulo $r\Omega$) using our generalized formula; the accretion rate $\mdot$.  The {\it black curve} shows $\mdot$ with Bai's formula (always $+$ve), and the {\it colored curves} show $\mdot$ with our formula ({\it red}: positive (inward) $\mdot$, {\it blue}: negative (outward) $\mdot$). Our $\mdot$ are overwhelmingly $+$ve for this dust-to-gas ratio, but not for others (following plots). }
\end{figure}

\begin{figure}
\includegraphics[scale=0.7, angle=90]{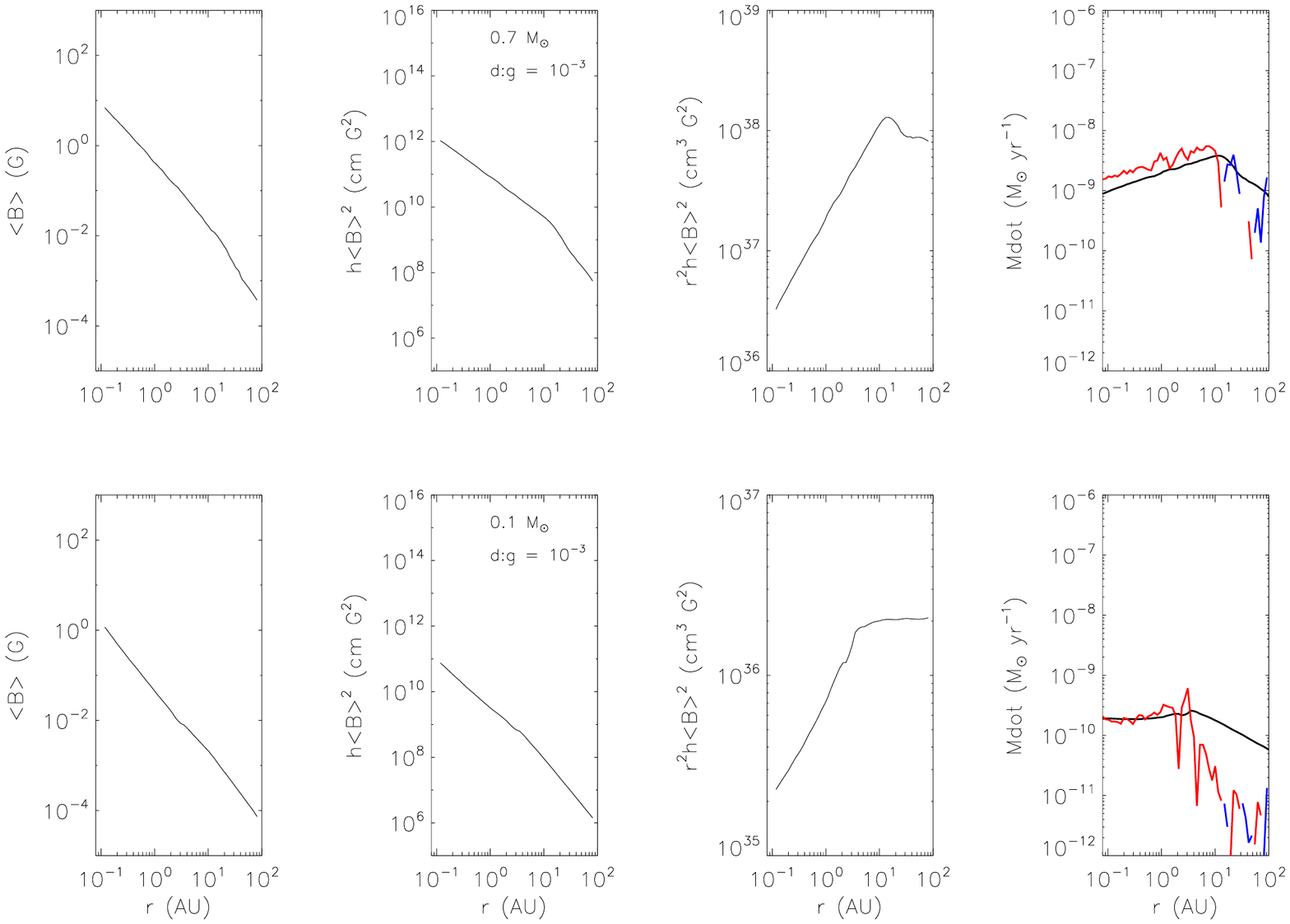}
\caption{Same as Fig.\,13, but with dust-to-gas ratio = $10^{-3}$.}
\end{figure}

\begin{figure}
\includegraphics[scale=0.7, angle=90]{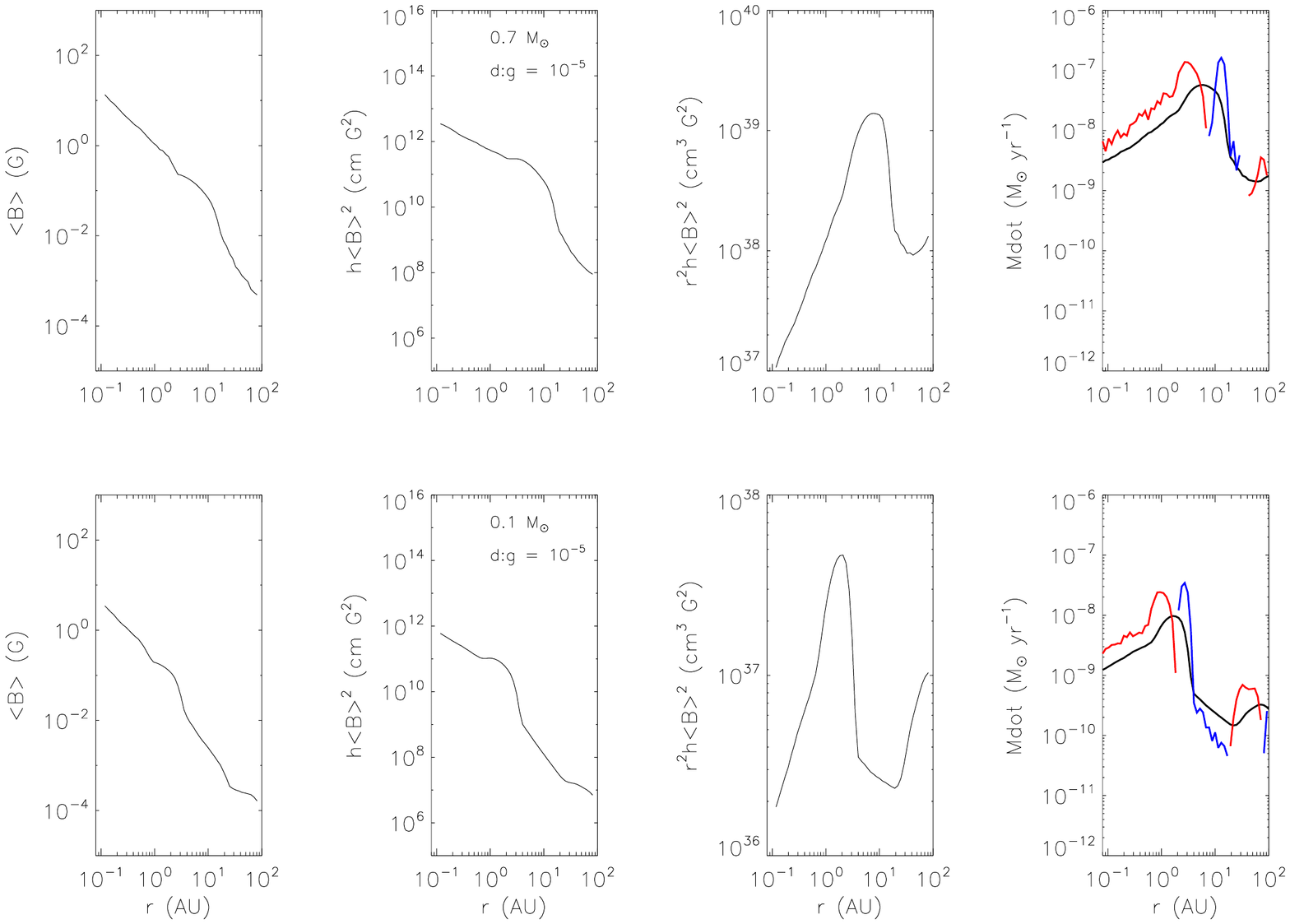}
\caption{Same as Fig.\,13, but with dust-to-gas ratio = $10^{-5}$.}
\end{figure}

\begin{figure}
\includegraphics[scale=0.7, angle=90]{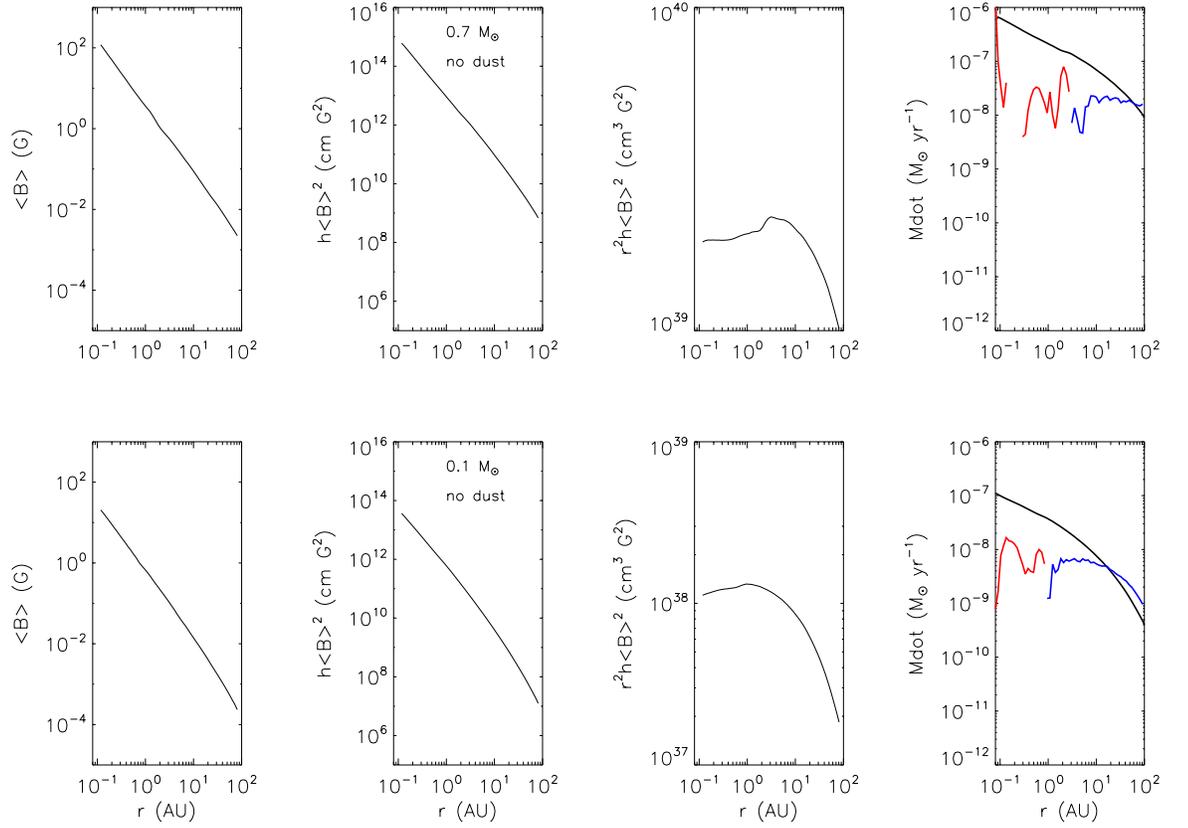}
\caption{Same as Fig.\,13, but with dust-to-gas ratio = 0 (no dust).}
\end{figure}

\end{document}